\documentclass[prb,aps,twocolumn,superscriptaddress,amssymb,floatfix,showpacs,amsfonts]{revtex4} 
\usepackage{graphicx,psfrag,amsmath,amssymb,float}
\usepackage{epstopdf}
\usepackage{color}
\input{epsf}
\usepackage{graphicx,natbib}

\newcommand{\bc}{\begin{center}}
\newcommand{\ec}{\end{center}}
\newcommand{\be}{\begin{equation}}
\newcommand{\ee}{\end{equation}}
\newcommand{\bea}{\begin{eqnarray}}
\newcommand{\eea}{\end{eqnarray}}

\def\12{\frac{1}{2}}

\makeatletter
\usepackage{psfrag}

\begin{document}

\title{Kondo Screening Cloud and the Charge Staircase in One-Dimensional Mesoscopic Devices}

\author{Rodrigo G. Pereira}

\affiliation{Department of Physics and Astronomy, University of British Columbia,
Vancouver, BC, Canada, V6T 1Z1}

\author{Nicolas Laflorencie}

\affiliation{Department of Physics and Astronomy, University of British Columbia,
Vancouver, BC, Canada, V6T 1Z1}
\affiliation{Insitute of Theoretical Physics, \'Ecole Polytechnique F\'ed\'erale de Lausanne, Switzerland}
\affiliation{Laboratoire de Physique des Solides, 
Universit\'e Paris-Sud, UMR-8502 CNRS, 91405 Orsay, France}
\author{Ian Affleck}
\affiliation{Department of Physics and Astronomy, University of British Columbia,
Vancouver, BC, Canada, V6T 1Z1}

\author{Bertrand~I.~Halperin}
\affiliation{Physics Department, Harvard University, Cambridge, MA
02138, USA} 

\date{\today}

\begin{abstract}
We propose that the finite size of the Kondo screening cloud, $\xi_{K}$,
can be probed by measuring the charge quantization in a one-dimensional
system coupled to a small quantum dot. When the chemical potential, $\mu$ 
in the system is varied at zero temperature, one should observe charge
steps whose locations are at values of $\mu$ that are controlled by the Kondo effect when the system size
$L$ is comparable to $\xi_{K}$.  We show that, 
if the standard Kondo model is used, the ratio between the widths
of the Coulomb blockade valleys with odd or even number of electrons 
is a universal scaling function of $\xi_{K}/L$. If we take into account electron-electron interactions in a single-channel wire, this ratio also depends on the parameters of the effective Luttinger model; in addition, the scaling is weakly violated by a marginal bulk interaction. 
For the geometry of a quantum dot
embedded in a ring, we show that the dependence of the charge steps on a magnetic flux through the ring is controlled by the size of the Kondo
screening cloud.
\end{abstract}


\maketitle
\section{Introduction}
The Kondo effect can be described as the strong renormalization of
the exchange coupling between an electron gas and a localized spin
at low energies.\cite{hewson} Below some characteristic energy scale,
known as the Kondo temperature $T_{K}$, a non-trivial many-body state
arises in which the localized spin forms a singlet with a conduction
electron. In the recent realizations of the Kondo effect,\cite{glazman}  
a semiconductor
quantum dot in the Coulomb blockade regime plays the role of a spin
$S=1/2$ impurity when the number of electrons in the dot is odd.
The Kondo temperature in these systems is a function of the coupling
between the dot and the leads and can be conveniently controlled by
gate voltages. As a clear signature of the Kondo effect, one observes
that the conductance through a quantum dot with $S=1/2$ is a universal
scaling function $G\left(T/T_{K}\right)$ and reaches the unitary
limit $G=2e^{2}/h$ when $T\ll T_{K}$ for symmetric coupling to the leads.\cite{delft} On the other
hand, one expects that in a finite system the infrared singularities
that give rise to the Kondo effect are cut off not by temperature,
but by the level spacing $\Delta$ if $\Delta\gg T$. For a one-dimensional
(1D) system with length $L$ and characteristic velocity $v$, the relevant
dimensionless parameter is $\Delta/T_{K}\sim\xi_{K}/L$, where $\xi_{K}\equiv v/T_K$
is identified with the size of the Kondo screening cloud, the mesoscopic
sized wave function of the electron that surrounds and screens the
localized spin. This large length scale ($\xi_{K}\sim0.1-1\mu\textrm{m}$ for typical values of $v$ and $T_K$)
is comparable with the size of currently studied mesoscopic devices,
which has motivated several proposals that the Kondo cloud should
manifest itself through finite size effects in such systems.\cite{kondobox,persistentcurrent,key-4,kaul} 

One familiar property of small metallic islands in the Coulomb blockade
regime is the quantization of charge. Even if the effects of electron-electron repulsion are neglected or subtracted off, each electron added to
the system costs a finite energy because the energy levels are discrete. Consequently, the number of electrons changes
by steps as one varies the chemical potential and sharp conductance
peaks are observed at the charge degeneracy points.\cite{glazman}
 In AlGaAs-GaAs heterostructures, the level
spacing of a 1D system with size $L\sim1\mu$m is of
order $\Delta\sim100\mu eV$, large enough to be resolved experimentally.
Clearly, the energy levels of the 1D wire should be affected by the Kondo interaction
with an adjacent dot. Therefore, the addition spectrum should exhibit
signatures of the finite size of the Kondo cloud. 

\psfrag{Vg}{$V_g$}

\psfrag{Vdw}{$V_{dw}$}

\psfrag{Phi}{$\Phi$}

\begin{figure}
\includegraphics[scale=0.5]{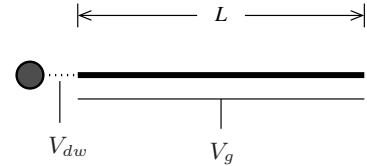}

\caption{Possible experimental setup. $V_{dw}$ controls the tunneling $t^{\prime}$ between
the small dot (on the left) and the wire and $V_{g}$ varies the chemical
potential in the wire. \label{cap:Possible-experimental-setup.}}
\end{figure}

We consider the geometry shown in Fig. \ref{cap:Possible-experimental-setup.}.
One of the ends of a wire of length $L$ with a large number of electrons
(such that $\Delta\ll\epsilon_{F}$, where $\epsilon_{F}$ is the
Fermi energy) is weakly coupled to a small quantum dot. The dot-wire tunneling is controlled by a gate voltage $V_{dw}$. Both the dot and the wire (which can be thought
of as a very long and thin dot) are in the Coulomb blockade regime.
The dot has a very large charging energy and a ground state
with $S=1/2$. We assume that the wire is very weakly connected to
a reservoir and that transfer of a single electron between wire
and reservoir could be measured by some Coulomb blockade technique.
This coupling is assumed weak enough so as not to affect the energy
levels of the wire-dot system. Experiments with small dots and large
dots have been performed, for example, in [\onlinecite{goldhaber-gordon}]
and considered theoretically in [\onlinecite{key-4,kaul}].  

Here we investigate the charge quantization in the wire-dot system using the basic Kondo model, ignoring 
charge fluctuations on the dot and assuming the wire contains 
a single channel. First we deal with the simplest case and ignore electron-electron interactions in the wire. In sections 
\ref{sec:wireweakcoup} and \ref{sec:wirestrongcoup}, we apply field theory methods to calculate the charge steps for the noninteracting 
case in both limits $L\ll\xi_K$ and $L\gg\xi_K$. In section \ref{sec:BA}, we compute the exact charge steps by solving numerically the 
Bethe Ansatz equations for the Kondo model. This approach  complements the field theory picture in the intermediate regime, $L\sim\xi_K$.
 The collapse of the numerical results confirms that the ratio of the widths of the charge plateaus with odd or even number of electrons is 
a universal scaling function of $\xi_{K}/L$. In section \ref{sec:luttinger}, we include short-range electron-electron interactions 
in the wire using the Luttinger model and discuss how the previous results are modified. In section \ref{sec:ring}, we consider 
the same problem for the geometry of a ring coupled to an embedded 
quantum dot, where the flux dependence of the charge step locations can also be studied. Section \ref{sec:conclusion} presents the conclusions.

\section{Weak coupling}\label{sec:wireweakcoup}

The tunneling between the wire and the small dot in the setup of Fig. \ref{cap:Possible-experimental-setup.} is well described
by the Anderson model:\begin{eqnarray}
H & = & \sum_{j=1}^{L-2}\left[-t\left(c_{j}^{\dagger}c_{j+1}^{\phantom{\dagger}}+h.c.\right)-\mu c_{j}^{\dagger}c_{j}^{\phantom{\dagger}}\right]+\varepsilon_{0}\sum_{\sigma}d_{\sigma}^{\dagger}d_{\sigma}^{\phantom{\dagger}}\nonumber \\
 &  & +Un_{d\uparrow}n_{d\downarrow}-t^{\prime}\sum_{\sigma}\left(d_{\sigma}^{\dagger}c_{1\sigma}^{\phantom{\dagger}}+h.c.\right).\label{eq:anderson}\end{eqnarray}
Here the wire is treated as a chain with $L-1$ sites and hopping
parameter $t$. The electron at site $j=1$ is coupled to a localized
state in the dot with tunneling amplitude $t^{\prime}$. The parameters
$\varepsilon_{0}$ and $U$ correspond to the energy and the Coulomb
repulsion for electrons in the dot, respectively. In the Coulomb blockade
regime $t^{\prime}\ll-\varepsilon_{0}\sim\varepsilon_{0}+U$, the
dot is singly occupied and the Anderson model is equivalent to the
Kondo model \cite{hewson}\begin{equation}
H=\sum_{j=1}^{L-2}\left[-t\left(c_{j}^{\dagger}c_{j+1}^{\phantom{\dagger}}+h.c.\right)-\mu c_{j}^{\dagger}c_{j}^{\phantom{\dagger}}\right]+Jc_{1}^{\dagger}\frac{\vec{\sigma}}{2}c_{1}^{\phantom{\dagger}}\cdot\vec{S},\end{equation}
where $\vec{S}=d^\dagger(\vec\sigma/2)d$ is the spin operator of the electron in the dot and
$J\sim t^{\prime2}/\left|\varepsilon_{0}\right|>0$ is the antiferromagnetic
Kondo coupling. In the following we assume that $J$ is independent of the chemical potential $\mu=eV_g$, which in practice  
may require that $\varepsilon_{0}$ and $t^{\prime}$ be tuned accordingly.

For $J=0$, the system reduces to an open chain and a free spin ($S=1/2$). In the continuum limit, we linearize the dispersion about the Fermi points and introduce  the  right- and left-moving components of the fermionic field $\Psi(x)$ for electrons in the wire\begin{equation}
\Psi\left(x\right)=e^{ik_{F}x}\psi_{R}\left(x\right)+e^{-ik_{F}x}\psi_{L}\left(x\right).\label{Psifield}\end{equation} 
The free Hamiltonian for the open chain in the continuum limit becomes \be
H_0  =  -iv_{F}\int_{0}^{L}dx\,\left(\psi_R^{\dagger}\partial_{x}\psi_R^{\phantom{\dagger}}+\psi_L^{\dagger}\partial_{x}\psi_L^{\phantom{\dagger}}\right).\label{H0wire}
\ee
The open boundary conditions $\Psi\left(0\right)=\Psi\left(L\right)=0$
imply that $\psi_{L,R}$ are not independent. Instead, $\psi_{L}$
can be regarded as the extension of $\psi_{R}$ to the negative-$x$
axis \begin{equation}\psi_{R}\left(-x\right)=-\psi_{L}\left(x\right).\label{boundcond}\end{equation}We can then
work with right movers only and write down an effective Kondo model $H=H_0+H_K$ in terms of $\psi_R$ only (we drop the index $R$ hereafter)\begin{equation}
H=-iv_{F}\int_{-L}^{L}dx\,\psi^{\dagger}\partial_{x}\psi+2\pi v_{F}\lambda_0\psi^{\dagger}\left(0\right)\frac{\vec{\sigma}}{2}\psi\left(0\right)\cdot\vec{S},
\label{Hamil}
\end{equation}
where $\lambda_0=2J\sin^2k_F/\pi v_F$ is the dimensionless Kondo
coupling.  For $\lambda_0\ll1$, the size of the Kondo screening
cloud (defined in the thermodynamic limit) is exponentially large: 
$\xi_{K}\sim(v_{F}/D)\, e^{1/\lambda_0}$, where $D\ll\epsilon_{F}$
is a high-energy cutoff. The boundary condition at the weak coupling
fixed point ($L/\xi_{K}\rightarrow0$) reads \be
\psi\left(-L\right)=e^{i2\delta}\psi\left(L\right),\label{bcweakcoup}\ee
where $\delta=k_{F}L$ mod $\pi$. Using the mode expansion \begin{equation}\psi\left(x\right)=-\frac{i}{\sqrt{2L}}\sum_{k}e^{ikx}c_{k},\label{psimodes}\end{equation}
we obtain the momentum eigenvalues $k_{n}=\left(n\pi-\delta\right)/L$, $n\in\mathbb{Z}$. 

\begin{figure}
\includegraphics[width=\columnwidth]{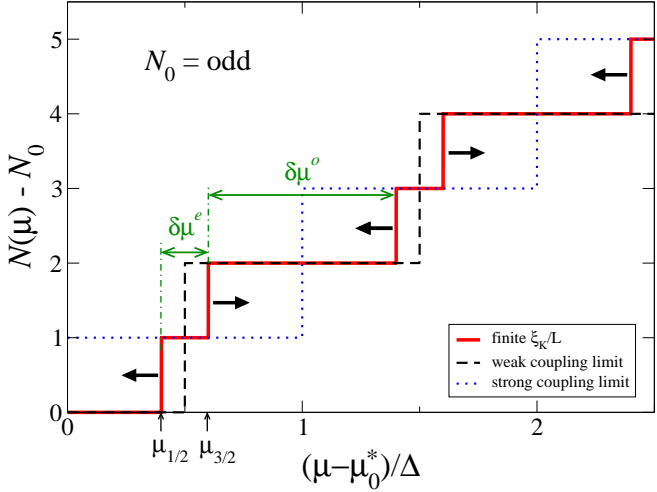}

\caption{(Color online) Charge quantization steps for the wire coupled to
a small quantum dot. The arrows indicate the direction in which the
single steps move as $\lambda\left(L\right)$ grows.\label{cap:Charge-steps.}}
\end{figure}

We denote by $N(\mu )$ the total number of electrons in the system, including 
the one in the quantum dot. We are interested in the elementary steps of $N$ around some initial value
$N_{0}\equiv N\left(\mu_{0}^*\right)$. At $T=0$, we calculate $N\left(\mu\right)$ as the integer part of $N$ that minimizes the thermodynamic potential
$\Omega\left(N\right)=E\left(N\right)-\mu N$, where $E\left(N\right)$
is the ground state energy. Defining $\mu_{\ell+\frac{1}{2}}$ as the value of $\mu$ where $N$ changes from $N_0+\ell$ to $N_0+\ell+1$, it follows  from the charge degeneracy condition $\Omega\left(N_{0}+\ell+1\right)=\Omega\left(N_{0}+\ell\right)$ that \begin{equation}\mu_{\ell+\frac{1}{2}}=E\left(N_{0}+\ell+1\right)-E\left(N_{0}+\ell\right).\label{mulformula}\end{equation}
For $\lambda_0=0$, we set $\mu_{0}^*$ halfway between the highest occupied
and the lowest unoccupied energy level of the open chain. In this
case, $N_{0}=$ odd: the ground state is doubly degenerate (total
spin $S_{tot}=1/2$) and consists of one electron in the dot and pairs
of conduction electrons in the Fermi sea. The energy levels measured
from $\mu_{0}^*$ are \be
\epsilon_{n}=\left(n-1/2\right)\Delta,\label{levelsopenchain}
\ee
where $\Delta=\pi v_{F}/L$ and $n$ is an integer. The energy of the ground state for $N=N_0+\ell$ electrons to $O(1/L)$ is 
\bea
E(N_0+\ell)&=&E_0+2\sum_{n=1}^{\ell/2}(\epsilon_n+\mu_0^*), \quad(\ell \textrm{ even})\nonumber\\
 &=&E_0+2\sum_{n=1}^{(\ell-1)/2}(\epsilon_n+\mu_0^*)\nonumber\\
& &+\epsilon_{(\ell+1)/2}+\mu_0^*, \quad(\ell \textrm{ odd})\eea
where $E_0=E(N_0)$. Using Eq. (\ref{levelsopenchain}), we get\be
E(N_0+\ell; \lambda_0=0)=E_0+\Delta\left(\frac{\ell^2}{4}+s^2\right)+\mu_0^*\ell,\label{nonintE}
\ee where $s=0$ for $\ell$ even and $s=1/2$ for $\ell$ odd. It is easy to see
that, as we raise $\mu>\mu_{0}^*$, $N$ jumps whenever $\mu$ crosses
an energy level $\epsilon_n$. Moreover, there are only double steps due to the
spin degeneracy of the single-particle states, i.e. $\mu_{2m+\frac{1}{2}}=\mu_{2m+\frac{3}{2}}$.
As a result, $N\left(\mu\right)-N_{0}$ takes on only even
values and $N$ is always odd (see dashed line in Fig. \ref{cap:Charge-steps.}).

We obtain the charge steps in the weak coupling limit by calculating
the correction to the ground state energy using perturbation theory
in $\lambda_0$ [\onlinecite{persistentcurrent}]. In the limit $\lambda_0\rightarrow0$,
the ground state for $N$ odd is still doubly degenerate and we write\be\left|GS\left(N=\textrm{odd}\right)\right\rangle =\prod_{n\leq0}c_{k_{n}}^{\uparrow\dagger}c_{k_{n}}^{\downarrow\dagger}\left|0\right\rangle \otimes\left|\gamma\right\rangle, \ee
for the values of $k_n$ defined below Eq. (\ref{psimodes}). The spin state of the dot is $\left|\gamma\right\rangle =\left|\Uparrow\right\rangle ,\left|\Downarrow\right\rangle$. For $N$ even, there
is one single electron occupying the state at the Fermi surface (with
momentum $k_{F}$). To lowest order in degenerate perturbation theory,
the ground state for $\lambda_0\rightarrow0$ is \be\left|GS\left(N=\textrm{even}\right)\right\rangle =\prod_{n\leq0}c_{k_{n}}^{\uparrow\dagger}c_{k_{n}}^{\downarrow\dagger}\left|0\right\rangle \otimes\left|s\right\rangle,\ee
where $\left|s\right\rangle \equiv[\left|k_{F}\uparrow\right\rangle \otimes\left|\Downarrow\right\rangle -\left|k_{F}\downarrow\right\rangle \otimes\left|\Uparrow\right\rangle ]/\sqrt{2}$
is the singlet state between the spin of the dot and the electron
at $k_{F}$. This state has $S_{tot}=0$ and $\left\langle \right.\vec{S}\cdot\vec{S}_{el}\left.\right\rangle =-3/4$,
where $\vec{S}_{el}\equiv\sum_{k}c_{k}^{\dagger}\frac{\vec{\sigma}}{2}c_{k}^{\phantom{\dagger}}$
is the total spin of the conduction electrons. Since the Kondo effect only involves the spin sector, the ground state energy must assume the general form \begin{equation}
E(N_0+\ell; \lambda)=E_0+\frac{\Delta}{4}\left[\ell^2+ f_s(\lambda_0,L)\right]+\mu_0^*\ell.
\label{GSenergy}\end{equation}According to Eq. (\ref{nonintE}), for $\lambda_0=0$, $f_0=0$ and $f_{1/2}=1$. For $\lambda_0\neq0$, the singlet formation
lowers $E\left(N\right)$ for $N=$ even relatively to $N=$ odd.
As a result, the Kondo interaction splits the double steps and gives
rise to small plateaus with $N$ even (Fig. \ref{cap:Charge-steps.}).
To $O\left(\lambda_0^{2}\right)$, we find
\begin{eqnarray}f(\lambda_0,L)&\equiv &f_{1/2}(\lambda_0,L)-f_{0}(\lambda_0,L)\nonumber \\ &=&1-3\left(\lambda_0+\lambda_0^2\ln \frac{DL}{v_F}+\dots\right)\label{f_weak},
\end{eqnarray}
Note the logarithmic divergence at $O\left(\lambda_0^{2}\right)$ as $L\rightarrow\infty$,
characteristic of the Kondo effect. We recognize the expansion of
the effective coupling $\lambda\left(L\right)\sim\left[\ln\left(\xi_{K}/L\right)\right]^{-1}$
in powers of the bare $\lambda_0$, as expected from scaling arguments.
We have verified that the scaling holds up to $O\left(\lambda_0^{3}\right)$,
\emph{i.e}., the function $f$ has no dependence on the cutoff $D$ but
the implicit one in the expansion of $\lambda\left(L\right)$.
This is remarkable given that $E\left(N\right)$ itself is cutoff
dependent. Based on this, we conjecture that $f$ is a universal scaling function
of $\xi_{K}/L$. It follows from Eqs. (\ref{mulformula}), (\ref{GSenergy}) and (\ref{f_weak}) that the charge steps occur at
\begin{equation}
\frac{\mu_{\ell+\frac{1}{2}}-\mu_0^*}{\Delta}=\left\lfloor\frac{\ell}{2}\right\rfloor+\frac{1}{2}-\frac{3}{4}(-1)^\ell \lambda(L),\label{eq:stepsweakcoup}\end{equation}
where $\left\lfloor x\right\rfloor$ is the floor function or the integer part of $x$, so that\bea
\left\lfloor\ell/2\right\rfloor &=&\left\{ \begin{array}{c}
\ell/2\quad,\quad\ell \textrm{ even}  \\
(\ell-1)/2\quad,\quad\ell \textrm{ odd}.  \end{array} \right.
\eea  We define the ratio (see Fig. \ref{cap:Charge-steps.})\begin{eqnarray}
R &\equiv&\frac{\delta\mu^{\rm{o}}}{\delta\mu^{\rm{e}}}\nonumber\\&=&\frac{E(N_0+3)-2 E(N_0+2)+E(N_0+1)}{E(N_0+2)-2 E(N_0+1)+E(N_0)} \nonumber \\ 
&=& \frac{1+f(\xi_K/L)}{1-f(\xi_K/L)}.\end{eqnarray}
From Eq. (\ref{f_weak}), we have that in the weak coupling
limit
\be
f(\xi_K/L\gg1)\approx 1-3[\ln(\xi_K/L)]^{-1},
\ee
and the ratio between the width of odd and even steps is
\begin{equation}
R\left(\frac{\xi_{K}}{L}\gg1\right)\approx\frac{2}{3}\ln\left(\frac{\xi_{K}}{L}\right)-\frac{1}{3}\ln \ln\left(\frac{\xi_K}{L}\right)+\textrm{const},
\label{WC}
\end{equation}
where we included the subleading $\ln\ln$ term in the expression of the effective Kondo coupling at scale $L$. A similar scaling function was found in [\onlinecite{kaul}] for the singlet-triplet gap (for fixed $N$) in the weak coupling limit.

\section{Strong coupling} \label{sec:wirestrongcoup}
When $\lambda\rightarrow\infty$,
the spin of the dot forms a singlet with one conduction electron and
decouples from the other electrons in the chain. In this limit the
Kondo cloud is small ($\xi_{K}/L\rightarrow0$). The strong coupling
boundary conditions reflect the $\pi/2$ phase shift for the particle-hole
symmetric case \begin{equation}
\psi\left(-L\right)=-e^{i2\delta}\psi\left(L\right).\end{equation} 
Note the minus sign relative to Eq. (\ref{bcweakcoup}).
This implies that the $k$ eigenvalues are shifted with respect to
weak coupling: $k_{n}=(n\pi+\pi/2-\delta)/L$. At the strong coupling fixed point
we recover the spin degeneracy of the single-particle states. The shifted energy levels are\be
\epsilon_{n}=n\Delta,\ee as measured from the original $\mu_{0}^*$. The ground state energy for $N=N_0+\ell$ is\be
E(N_0+\ell)=E(N_0)+\Delta\left[\frac{\ell^2}{4}+\left(s-\frac{1}{2}\right)^2\right]+\mu_0^*\ell,
\ee
where again $s=0$ for $\ell$ even and $s=1/2$ for $\ell$ odd.
The ground state consists now of a singlet
plus pairs of electrons in the wire, thus $N\left(\mu\right)$ is
always even. Fig. \ref{cap:Charge-steps.} illustrates how the charge
staircase evolves monotonically from weak to strong coupling.

We explore the limit $k_{F}^{-1}\ll\xi_{K}\ll L$ by working out a
local Fermi liquid theory.\cite{nozieres} The idea is that virtual
transitions of the singlet at $x=0$ induce a local interaction in
the spin sector of the conduction electrons. The leading irrelevant
operator that perturbs the strong coupling fixed point and respects
SU(2) symmetry is \begin{equation}
H_{FL}=-\frac{2\pi^{2}}{3}\frac{v_{F}^{2}}{T_{K}}\left[\psi^{\dagger}\left(0\right)\frac{\vec{\sigma}}{2}\psi\left(0\right)\right]^{2},\label{eq:localFLinteraction}\end{equation}
 where the prefactor is fixed such that the impurity susceptibility
is $\chi_{imp}=1/(4T_{K})$. This interaction lowers the ground state
energy when there is an odd number of remaining conduction electrons
($S_{el}=1/2$, or $s=0$), thus splitting the charge steps in the strong coupling
limit. If we introduce the function $f$ as in (\ref{GSenergy}) and (\ref{f_weak}), we find to lowest order in $1/T_{K}$\begin{equation} f(\xi_K/L)=-1+\frac{\pi\xi_K}{2L}+O\left((\xi_K/L)^2\right).
\end{equation}
The charge steps now appear at
\begin{equation}
\frac{\mu_{\ell+\frac{1}{2}}-\mu_0^*}{\Delta}=\left\lfloor\frac{\ell+1}{2}\right\rfloor+(-1)^\ell\frac{\pi}{8}\frac{\xi_{K}}{L},\end{equation}
from which we obtain \begin{equation}
R\left(\frac{\xi_{K}}{L}\ll1\right)\approx\frac{\pi}{4}\frac{\xi_{K}}{L}.
\label{SC}
\end{equation}

At this point we would like to comment about the effect of particle-hole symmetry breaking for this wire-dot geometry. For the Anderson model of Eq. (\ref{eq:anderson}), particle-hole symmetry is absent if the system is away from half-filling.  We can account for this by adding to the
Hamiltonian a scattering potential term \be
H_p=2\pi v_F V\psi^\dagger(0)\psi^{\phantom\dagger}(0),
\ee
where $V$ is of order the bare Kondo coupling $\lambda_0\ll1$.\cite{hewson} This term is strictly marginal and can be treat by  first-order perturbation theory in both weak and strong coupling limits. Its effect is simply to shift the position of the charge steps by \be\mu_{\ell+\frac{1}{2}}\to\mu_{\ell+\frac{1}{2}}+V\Delta.\ee
Since the shift is independent of the parity of $\ell$, the potential scattering term does not lift the spin degeneracy of the charge steps and has  no effect on the ratio $R=\delta\mu^o/\mu^e$.

\section{Bethe Ansatz results} \label{sec:BA}

\begin{figure}
\includegraphics[width=\columnwidth]{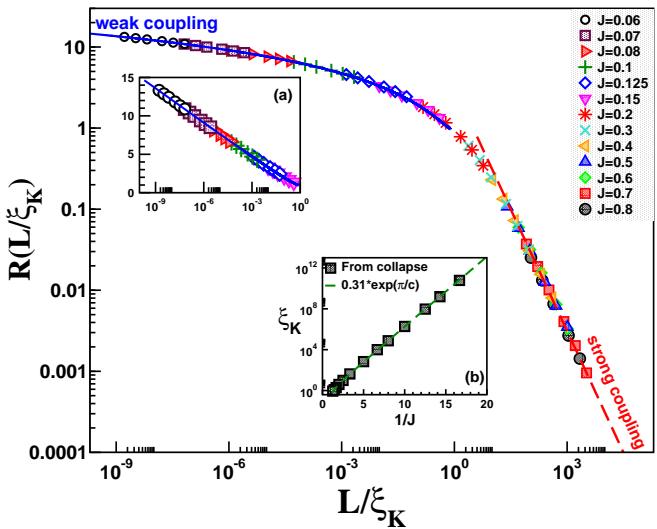}
\caption{(Color online) Universal ratio $R=\delta\mu^{\rm{o}}/\delta\mu^{\rm{e}}$ as a scaling
function of $\xi_{K}/L$. Bethe ansatz results obtained for various systems sizes ($N_0=51,~101,~201,~501,~1001,~2001$) and 13 different values of the Kondo exchange $J$, indicated by different symbols. For each value of $J$, the system lengths $L$ have been rescaled $L\to L/\xi_K(J)$ in order to obtain the best collapse of the data using the strong coupling curve Eq.~(\ref{SC}) (dashed red line) as a support for the rest of the collapse. 
The weak coupling regime for $L\ll\xi_K$, enlarged in the inset (a), is described by the weak coupling expansion Eq.~(\ref{WC}) with ${\rm{constant}}\simeq 0.33$ (continuous blue curve).
Inset (b): The Kondo length scale, extracted from the universal data collapse of the main panel (black squares), is described by the exponential fit Eq.~(\ref{xi}) with $\xi_0\simeq 0.31$ (dashed green line).}
\label{BA}
\end{figure}
In order to calculate the scaling property of $R=\delta\mu^{\rm{o}}/\delta\mu^{\rm{e}}$ for any $L/\xi_K$ we use the Bethe ansatz (BA) solution of the one-channel Kondo problem.\cite{Andrei80} We start with a half-filled
band of $N_0-1$ conduction electrons ($N_0$ odd) coupled to a localized impurity spin, corresponding 
to a system size $L=(N_0-1)/2$ in units where the Bethe ansatz cut-off parameter $D$, 
related to the bandwidth, is set equal to one (see [\onlinecite{Andrei80}]). 
Since the BA solution can be obtained for any filling factor, we can add particles one by one to the system ($N=N_{0},~N_{0}+1,~N_{0}+2,\hdots$) and compute the corresponding energies. This has been done  by solving numerically the coupled BA equations,\cite{Andrei80} using a standard Newton-Rapson method.
In Fig.~\ref{BA} we present results obtained for the ratio 
$R=\delta\mu^{\rm{o}}/\delta\mu^{\rm{e}}$ with $N_0=51,~101,~201,~501,~1001,~2001$ and 13 different values of the Kondo coupling in the range $0.06\le J\le 0.8$.
The universal scaling curve has been obtained by rescaling the $x$-axis $L\to L/\xi_K(J)$ in order to get the best collapse of the data. The entire crossover curve is obtained, ranging from the strong coupling $L\gg\xi_K$  to the weak coupling regime $L\ll\xi_K$. Both strong coupling [Eq.~\ref{SC})] and weak coupling [Eq.~(\ref{WC})] results are perfectly reproduced. As displayed in the inset (a) of Fig.~\ref{BA}, BA results agree with Eq.~(\ref{WC}) using only one fitting parameter: const$\simeq 0.33$. The  Kondo length scale, shown in the inset (b) of Fig.~\ref{BA}, displays the expected exponential behavior\cite{Andrei80}
\be
\xi_K(J)=\xi_0{\rm{e}}^{\pi/c},
\label{xi}
\ee
with $c=2J/(1-3J^2/4)$. Fitting $\xi_K$ to this expression gives $\xi_0\simeq 0.31$ (see inset (b) of Fig.~\ref{BA}) which is in very good agreement with the expected value of $\xi_0=1/\pi$, resulting\cite{hewson} from the impurity susceptibility normalized to $1/(4T_K)$.

\section{Luttinger liquid effects} \label{sec:luttinger}
Let us now include the effect of electron-electron interactions in the wire. We consider the Hamiltonian\cite{fabrizio} \be
H=H_0+H_{I}+H_K.
\ee
The kinetic energy part $H_0$ is given, in the low energy limit, by Eq. (\ref{H0wire}). The short-range  interactions are described by\be
H_{I}=\frac{1}{2}\sum_{\sigma\sigma^\prime}\int dx\int dx^\prime \,\rho_\sigma(x)V(x-x^\prime)\rho_{\sigma^\prime}(x^\prime),
\ee
where $V(x)$ is the screened Coulomb potential and $\rho_\sigma(x)\equiv \Psi^\dagger_\sigma(x)\Psi^{\phantom\dagger}_{\sigma}(x)$ is the density of electrons with spin $\sigma$. In the low-energy limit, we expand $\Psi(x)$ in terms of right and left movers as in Eq. (\ref{Psifield}) and write\bea
\rho_{\sigma}(x)&=&\rho_{R,\sigma}(x)+\rho_{L,\sigma}(x)\nonumber\\&&+[e^{i2k_Fx}\psi_{L\,\sigma}^\dagger(x)\psi^{\phantom\dagger}_{R\,\sigma}(x)+h.c.],
\eea
where $\psi_{R/L\,\sigma}(x)\equiv\psi_{R/L\,\sigma}^\dagger(x)\psi^{\phantom\dagger}_{R/L\,\sigma}(x)$. Separating the processes that involve small momentum transfer ($q\approx0$) from the backscattering ones ($q\approx2k_F$), we obtain in the conventional g-ology notation
\bea
H_I&=&\sum_\sigma\int_0^L dx\left\{\frac{g_{4\parallel}}{2}\left[\rho_{R,\sigma}^2+\rho_{L,\sigma}^2\right]\right.\nonumber\\
&&+g_{2\parallel}\rho_{R,\sigma}\rho_{L,\sigma}+\frac{g_{4\perp}}{2}\left[\rho_{R,\sigma}\rho_{R,-\sigma}+\rho_{L,\sigma}\rho_{L,-\sigma}\right]\nonumber\\&&\left.+g_{2\perp}\rho_{R,\sigma}\rho_{L,-\sigma}\right\}+H_{ex},
\eea
where $g_{2\parallel}=g_{2\perp}=g_{4\parallel}=g_{4\perp}= \tilde{V}(0)$ for a spin-independent interaction potential, with $\tilde{V}(0)$ the Fourier transform of $V(x)$ at momentum $q=0$. 
The term $H_{ex}$ contains the backscattering interaction processes. As we will discuss below, this interaction is marginally irrelevant in the sense of the renormalization group. If we  neglect $H_{ex}$ for a moment, the resulting Hamiltonian is exactly solvable by bosonization.\cite{haldane81} For open boundary conditions, we can work with right movers on a ring with size $2L$ and the boundary conditions of Eq. (\ref{boundcond}). After introducing charge and spin densities \be
\rho_{R,c/s}=\frac{\rho_{R,\uparrow}\pm\rho_{R,\downarrow}}{\sqrt{2}},
\ee 
one finds that the charge and spin degrees of freedom decouple. The effective model for electrons in the wire with zero Kondo coupling is the Luttinger model with open boundary conditions\cite{fabrizio}
\be
H_0+H_I=H_{LL}+H_{ex},\label{luttingermodel}
\ee
where (setting $\mu_0^*=0$)
\be
H_{LL}=\frac{\pi v_c}{4K_cL}\hat N^2+\frac{\pi v_s}{L}(\hat S^z)^2+\sum_{\scriptsize\begin{array}{c}q>0
\\ \nu=c,s \end{array}} v_\nu |q|b^\dagger_{q\nu}b^{\phantom\dagger}_{q\nu},\label{bosonmodes}
\ee
where $\hat N$ and $\hat S^z$ are the zero modes for charge and spin excitations, $v_c$ and $v_s$ are the velocities of the collective charge and spin modes created by the bosonic operators $b^\dagger_{q c/s}$, and $K_c$ is the Luttinger parameter for the charge sector. We assume spin SU(2) symmetry and set the Luttinger parameter for the spin sector to be $K_s=1$.  From Eq. (\ref{bosonmodes}), we have that the energy of the ground state for $\ell$ extra electrons (neglecting $H_{ex}$) is \begin{equation} 
E(N_0+\ell,\lambda_0=0)=E_0+\frac{\pi v_c}{4 K_cL}\, \ell^2 +\frac{\pi v_s}{L} \,s^2,\label{E_inter}\end{equation}
The noninteracting case in Eq. (\ref{nonintE}) corresponds to $v_c=v_s=v_F$ and $K_c=1$. For repulsive interactions, $K_c<1$ and $v_c>v_s$. For weak interactions, we have the perturbative results from bosonization\be K_c=\frac{v_F}{v_c}\approx\left[1+\frac{2\tilde{V}(0)}{\pi v_F}\right]^{-1/2} \ee and $v_s=v_F$.

The term $H_{ex}$ on the rhs of Eq. (\ref{luttingermodel}) is sometimes called the marginally irrelevant bulk interaction. This is entirely 
in the spin sector and is analogous to the exchange interaction in the constant interaction model.\cite{glazman} 
The operator can be written as\begin{equation}
H_{ex}=-2\pi g_0v_s\int dx\, \vec{J}_L(x)\cdot \vec{J}_R(x),\label{mibi}
\end{equation}
where $\vec{J}_{L/R}\equiv \psi_{L/R}^\dagger(\vec{\sigma}/2)\psi_{L/R}^{\phantom{\dagger}}$.
The bare coupling constant $g_0$ is proportional to $\tilde{V}(2k_F)$ (backward scattering process).\cite{Schultz} It is typically very small if the screening length is much larger than $k_F^{-1}$. The renormalized coupling $g(L)$ obeys the renormalization group (RG) equation\begin{equation}
\frac{dg}{dl}=-g^2-\frac{g^3}{2}+\dots,
\end{equation} where $l=\ln (DL/v_s)$. Precisely the same interaction appears in a spin chain (see, for example, [\onlinecite{Qin}]). The solution to $O(g^2)$ is \begin{equation}
g(l)\approx \frac{g_0}{1+g_0l}.
\end{equation} Therefore, $g(L)\sim 1/\ln(L)$ vanishes in the limit $L\rightarrow \infty$. The correction to the energy of the lowest energy state due to $H_{ex}$ is\cite{Qin}
\begin{equation}
\delta E(N_0+\ell,\lambda_0=0)=-\frac{3\pi v_s}{2L}g(L)\, s^2+O(g^2).
\end{equation}

The Kondo effect has also been studied in Luttinger liquids.\cite{Furusaki} One crucial point is that the impurity spin couples only to the spin degrees of freedom of the Luttinger liquid \emph{when the impurity spin is at one end}. This is not true if the spin couples far from the end. 
The Kondo interaction is as before\be
H_K=J\Psi^\dagger(\varepsilon)\frac{\vec{\sigma}}{2}\Psi(\varepsilon)\cdot \vec{S},
\ee
where $\varepsilon\sim k_F^{-1}\ll L$ ($\varepsilon=1$ in the lattice model). In general, the spin density operator is \begin{equation}
\Psi^\dagger(x)\frac{\vec{\sigma}}{2}\Psi(x)\sim \vec{J}_L+\vec{J}_R+\left(e^{i2k_Fx}\psi_L^\dagger\frac{\vec{\sigma}}{2}\psi_R^{\phantom{\dagger}}+h.c.\right).
\end{equation}
When we bosonize, the left and right spin densities $\vec{J}_{L/R}$ can be expressed entirely in terms of the spin boson, but the
 $2k_F$ terms become products of spin and charge operators. However, for open boundary conditions we 
can use Eq. (\ref{boundcond}) and the entire spin density at $x=0$ becomes proportional to $\vec{J}_{R}(0)$. 
For this reason, spin-charge separation is preserved (up to irrelevant operators) for the geometry of Fig. 1 
with a finite Kondo interaction.  The bulk Kondo interations only important effects on the spin sector, 
in the low energy limit, are to 
modify somewhat the spin velocity, $v_s$ and to introduce the marginally irrelevant bulk 
interaction of Eq. (\ref{mibi}).
The finite size energies still break up into a sum of spin and charge parts. 
The charge part is exactly as in Eq. (\ref{E_inter}). In the spin part, the dimensionless Kondo coupling is 
defined as $\lambda_0=2J\sin^2(k_F\varepsilon)/\pi v_s$, with $v_s$ replacing $v_F$ in contrast with Eq. (\ref{Hamil}). 
The Kondo coupling is still marginally relevant for $K_s=1$. We can then write 
\begin{equation}E(N_0+\ell,\lambda_0)=\frac{\pi v_c}{4 K_cL}\, \ell^2 +\frac{\pi v_s}{4L}f_s(\lambda,g,L).
\end{equation}
The scaling functions $f_s(x)$, $s=0,1/2$, are the same ones that we calculated ignoring Luttinger liquid interactions,
 apart from the contribution from the marginally irrelevant bulk interaction. To lowest order in $\lambda$ and $g$, we have
 \begin{eqnarray}
f(\lambda,g,L)&=&f_{1/2}(\lambda,g,L)-f_0(\lambda,g,L)\nonumber \\
&=&1-\frac{3}{2}g(L)-3\lambda(L).\label{f_intweak}
\end{eqnarray}
Note that both the bulk marginal operator and the Kondo interaction reduce the energy when $N$ is even.

The RG equation for the Kondo coupling $\lambda$ is modified by the marginal bulk operator\cite{Laflorencie}\begin{equation}
\frac{d\lambda}{dl}=\lambda^2+g\lambda+\dots.\label{RGKondo}
\end{equation}
Solving this equation in the presence of the second term, one finds\begin{equation}
\lambda(L)=\frac{2\ln(L/L_1)}{\ln^2(\xi_K/L_1)-\ln^2(L/L_1)+2\ln(\xi_K/L_1)},
\end{equation}
where $\xi_K$ is defined such that $\lambda(L=\xi_K)=1$, and $L_1$ is a characteristic length scale for the bulk marginal interaction defined by $g(L)\approx 1/\ln(L/L_1)$. If the bare $g_0$ is small, then $L_1\sim (v_s/D)e^{-1/g_0} \ll v_s/D$. In the limit $L_1\rightarrow 0$ we recover the scaling function $\lambda=\lambda(\xi_K/L)\sim [\ln(\xi_K/L)]^{-1}$. In general, we expect $\lambda=\lambda(\xi_K/L, g(L))$.  Eq. (\ref{RGKondo}) also implies an unusual dependence of $\xi_K$ on the bare Kondo coupling \cite{Laflorencie,FZ}\begin{equation}
\xi_K\sim L_1 \exp\left[-c+\sqrt{\frac{2}{\lambda}\ln\left(\frac{v_s}{DL_1}\right)+\ln^2\left(\frac{v_s}{DL_1}\right)+c^2}\right],\label{longxi_K}
\end{equation}
where $c$ is a positive constant of $O(1)$. In the limit $L_1\ll v_s/D$ we recover the usual result $\xi_K\sim (v_s/D)e^{-1/\lambda_0}$. In the limit $\lambda\rightarrow 0$ with $g$ held fixed this becomes $\xi_K\propto e^{\textrm{const}/\sqrt{\lambda_0}}$.

In the charge staircase (see again Fig. \ref{cap:Charge-steps.}), the width of the plateaus is normalized by the charge 
addition energy $\Delta_c=\pi v_c/K_c L$. The charge steps for $\xi_K\gg L$ are given by
\begin{equation}
\frac{\mu_{\ell+\frac{1}{2}}-\mu_0^*}{\Delta_c}=\left\lfloor\frac{\ell}{2}\right\rfloor+\frac{1}{2}-\frac{(-1)^\ell}{4} \left[1-u\, f\left(\frac{\xi_K}{L},g(L)\right)\right],\label{interac_steps}\end{equation}
where 
\be u\equiv v_sK_c/v_c<1\ee
 and $f$ is given by Eq. (\ref{f_intweak}). As the charge and spin addition energies are different for $u\neq 1$,\cite{Kleimann} 
there are no double steps in the interacting case even for $\lambda\rightarrow 0$. The ratio between odd and even steps including 
interaction is
\begin{equation}
\tilde{R}=\frac{1+u\,f(\xi_K/L,g(L))}{1-u\,f(\xi_K/L,g(L))}\label{Randf}.
\end{equation}
Thus we see that the main effect of the screened bulk Coulomb interactions is to suppress $\tilde{R}$ somewhat 
due to the reduction of the parameter $u$ from $1$. As $u\to 0$ (for very strong Coulomb interactions) 
the even-odd effect disappears entirely and we see simple Coulomb blockade type steps. For cleaved edge 
overgrowth quantum wires, $u\approx 0.5$.\cite{Auslaender2000,Auslaender2002,Steinberg2006} For carbon nanotubes,
one finds, typically, $u\approx 0.1$.\cite{Bockrath} (For further discussion of the experimental possibilities 
see Sec. VII.) 
For $\xi_K\gg L$,
\begin{equation}
\tilde{R}\approx\frac{1+u-3u\lambda(L)-\frac{3}{2}ug(L)}{1-u+3u\lambda(L)+\frac{3}{2}ug(L)}\label{R_weakcoup}.
\end{equation}

Now consider the strong coupling limit, $\xi_K\ll L$. The Fermi liquid interaction is the same as in Eq. (\ref{eq:localFLinteraction}), but with $\xi_K$ given by Eq. (\ref{longxi_K}). In contrast with Eq. (\ref{f_intweak}), the marginal operator now lowers the ground state energy when $N$ is odd,  because in the strong coupling limit the number of free electrons in the wire in $N-2$. Thus we have in this case 
\begin{equation}
f(\xi_K/L,g(L))=-1+\frac{3}{2}g(L)+\frac{\pi\xi_K}{2L}.\label{f_intstr}
\end{equation} 
The charge steps for $\xi_K\ll L$ are given by Eq. (\ref{interac_steps}) with $f$ taken from Eq. (\ref{f_intstr}). The ratio becomes \begin{equation}
\tilde{R}=\frac{1-u+\frac{3}{2}ug(L)+u\frac{\pi\xi_K}{2L}}{1+u-\frac{3}{2}ug(L)-u\frac{\pi\xi_K}{2L}}\label{R_strcoup}.
\end{equation}

If we set $g=0$, we can express the ratio $\tilde{R}$ in terms of the ratio calculated for the noninteracting case\begin{equation}
\tilde{R}=\frac{1-u+(1+u)R}{1+u+(1-u)R},
\end{equation}
where $R=R(\xi_K/L)$ is the scaling function shown in Fig. \ref{BA}. For $u\neq1$, the function $\tilde{R}$ is finite at both weak and strong coupling fixed points. In the weak coupling limit, $R\rightarrow \infty$ and $\tilde{R}\rightarrow (1+u)/(1-u)$. In the strong coupling limit, $R\rightarrow 0$ and $\tilde{R}\rightarrow (1-u)/(1+u)$.
\begin{figure}
\includegraphics[width=\columnwidth]{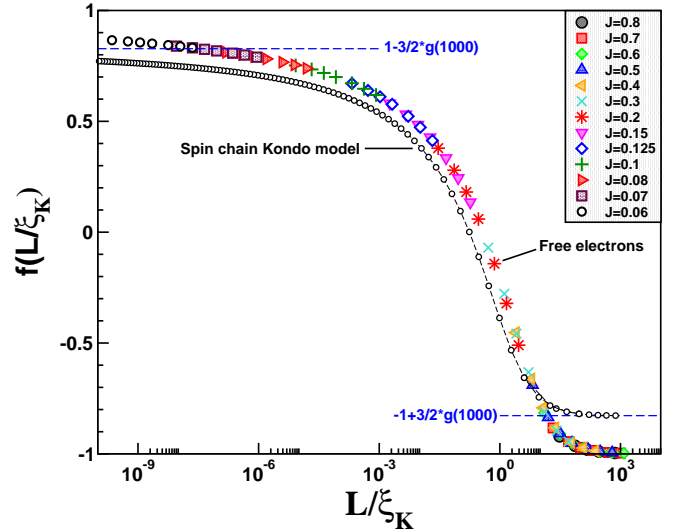}
\caption{(Color online) Effect of the marginally irrelevant bulk operator on the scaling function $f(\xi_K/L,g(L))$ calculated from the Bethe ansatz. The symbols correspond to the noninteracting case, $g(L)=0$, as in Fig. \ref{BA}. The black circles are obtained with the Bethe ansatz solution of  the spin chain Kondo model (\ref{eq:SCKM}) for $L=1000$ ($g(1000)\simeq 0.115$). }
\label{fgL}
\end{figure}

A nonzero $g_0$ leads to  a weak violation of scaling because  $f$ is not only a function of $\xi_K/L$ but also of  $g(L)$.  The effective $g(L)$ 
is typically small in quantum
wires, except at very low electron densities, and could in practice
be treated as a fitting parameter. We can write down a general expansion for $f$ in powers of $g$:
 \begin{equation}
f(\xi_K/L,g)=f_0(\xi_K/L)+g f_1(\xi_K/L)+O(g^2).
\end{equation}
It follows from Eqs. (\ref{f_intweak}) and (\ref{f_intstr}) that $f_1(\xi_K/L\gg 1)= -3/2+O(\lambda(L))$ and $f_1(\xi_K/L\ll1)= 3/2+O(\xi_K/L)$.

\begin{figure}[!ht]
\begin{center}
\includegraphics[width=7cm,clip]{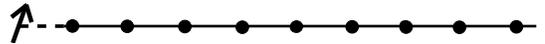}
\end{center}
\caption{Schematic picture for an open Heisenberg chain coupled to a spin impurity (arrow) at the left boundary via a 
weak antiferromagnetic exchange $J_K'$ (dashed bond). }
\label{fig:NNchain}
\end{figure}

In fact, the function $f(\xi_K/L,g)$ can be calculated in the framework of a Heisenberg spin chain model with a 
weak coupling $J'_K$ at the end of the chain~\cite{Laflorencie,FZ}
\be
{\cal H}=J_1\sum_{i=1}^{L-1}{\vec{S}}_{i}\cdot{\vec{S}}_{i+1}+{\cal H}_{imp},\ \ \ 
{\cal H}_{imp}=J_K' {\vec{S}}_{imp}\cdot{\vec{S}}_1.
\label{eq:SCKM}
\ee
This model is depicted in Fig.~\ref{fig:NNchain}.
Its low energy limit is the same as the spin sector of the Luttinger liquid.\cite{Eggert,FZ,Laflorencie}  In this case the bulk marginal 
coupling has a bare value, $g_0$, of $O(1)$. The Kondo coupling is proportional to $J_K'$.~\cite{Laflorencie} 
Since the charge sector separates from the spin sector anyway, in the Luttinger liquid Kondo model, 
it follows that the needed function $f(\xi_K/L,g(L))$ can be determined from the spin chain model. 

The function $f(\xi_K/L,g(L))$ can be evaluated via the energy 
difference between the ground states for even and odd chain length, of total spin 0 and 1/2:
\be
E_0 - E_{1/2} \to \frac{\pi v_s}{4L} f.
\ee
One can extract this quantity  {\it exactly}  using the BA solution~\cite{FZ} of the model (\ref{eq:SCKM}) by computing $\frac{1}{2}\left[E(L+2)+E(L)-2E(L+1)\right]$ with $L$ even for various Kondo coupling strengths $J'_{K}$. This has been achieved for $L=1000$ as shown in Fig.~\ref{fgL}, where one can see the full scaling function with the two limiting cases
\be
f\to\left\{
\begin{array}{rl}
1-\frac{3}{2}g & ~~\mathrm{if}\ {L}/{\xi_K} \to 0 \\
-1+\frac{3}{2}g & ~~\mathrm{if}\ {L}/{\xi_K} \to \infty.
\end{array}
\right.
\ee
In order to obtain the scaling function $f$ of the spin chain Kondo model in the entire regime of $L/\xi_K$ shown in Fig.~\ref{fgL} with $L=1000$, we had to convert $J'_K\to \xi_K$ using the unusual exponential square root behavior for the Kondo length scale (\ref{longxi_K}). In order to make things quantitative, we
used the following conversion~\cite{FZ} for $\xi_K(J'_K)$
\be \xi_K= \xi_0\exp\left(\pi \sqrt{1/J_K'-1}\right).\label{eq:xiKFZ}\ee
with~\cite{FZ,Laflorencie} $\xi_0=\frac{\sqrt{e/\pi}}{2}$.
It is reassuring to note that, even for this large value of bare coupling $g_0$,
 the effects of the bulk marginal operator 
on the cross-over function $f$ are fairly small for $L=1000$. We might expect these effects to 
be even smaller for the Luttinger model with realistic parameters chosen to describe the 
systems discussed in Sec. VII. 

\section{Flux dependence in quantum rings} \label{sec:ring}
\begin{figure}
\includegraphics[width=\columnwidth]{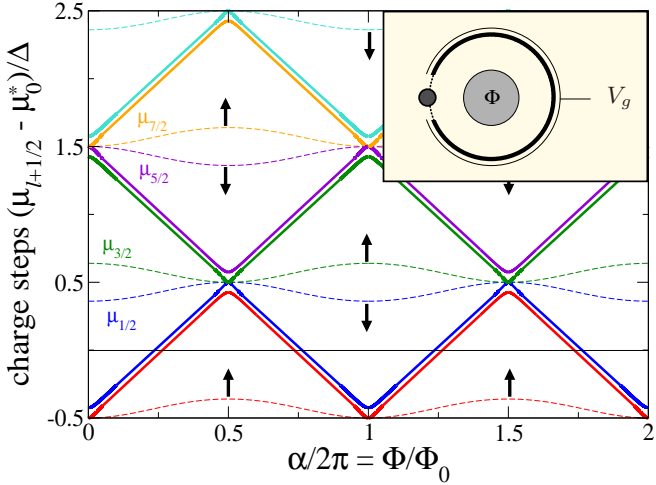}

\caption{(Color online) Flux dependence of the charge steps in the ring with
an embedded quantum dot (inset), assuming no potential scattering ($V=0$). Dashed lines: weak coupling
regime ($\xi_{K}\gg L$); solid lines: strong coupling regime
($\xi_{K}\ll L$).\label{cap:(Color-online)-Flux}}
\end{figure}

The effect of the Kondo screening cloud on the charge steps of a 1D system is not unique to the wire geometry. One interesting alternative is to suppose that the quantum dot is embedded in a ring with
circumference $L$ (inset of Fig. \ref{cap:(Color-online)-Flux}).
This geometry offers the possibility of looking at the dependence
of the charge steps on the magnetic flux $\Phi$ threading the ring, which is related to the persistent current 
for the embedded quantum dot.\cite{persistentcurrent,Zvyagin} In this section we focus again on the noninteracting case. 
 For simplicity, we assume that the coupling between the dot and the leads is parity symmetric. 
We can write the Hamiltonian (for zero flux) as $H=H_0+H_K$, where $H_0$ is the same as in Eq. (\ref{H0wire}) and the Kondo interaction is\begin{equation}H_K=J [\Psi^{\dagger}(\varepsilon)+\Psi^{\dagger}(L-\varepsilon)]\frac{\vec{\sigma}}{2}[\Psi(\varepsilon)+\Psi(L-\varepsilon)]\cdot \vec{S},\end{equation}where $\varepsilon\sim k_F^{-1}\ll L$. In the symmetric case it is convenient to label the eigenstate
of the open chain by their symmetry under the parity transformation
$x\rightarrow L-x$. Expanding $\Psi(x)$ as in Eq. (\ref{Psifield}), we have\bea
\Psi(\varepsilon)+\Psi(L-\varepsilon)=e^{ik_F\varepsilon}[\psi_R(\varepsilon)+e^{-ik_FL}\psi_L(L-\varepsilon)]\nonumber\\+e^{ik_F(L-\varepsilon)}[\psi_R(L-\varepsilon)+e^{-ik_FL}\psi_L(\varepsilon)].
\eea This motivates defining the even and odd fields \begin{equation}\psi_{e,o}\left(x\right)=\frac{\psi_{R}\left(x\right)\pm e^{-ik_FL}\psi_{L}\left(L-x\right)}{\sqrt{2}}.\end{equation}
This way,  for zero flux $H_K$ involves only the even channel since\be
\frac{\Psi(\varepsilon)+\Psi(L-\varepsilon)}{\sqrt{2}}=e^{ik_F\varepsilon}\psi_e(\epsilon)+e^{ik_F(L-\varepsilon)}\psi_e(L-\epsilon).
\ee

For $J=0$, the system is equivalent to an open chain and a free spin. In terms of even and odd fields, the free Hamiltonian of
 Eq. (\ref{H0wire}) is simply\cite{persistentcurrent} \begin{equation}
H_0  =  -iv_{F}\int_{0}^{L}dx\,\left(\psi_e^{\dagger}\partial_{x}\psi_e^{\phantom{\dagger}}+\psi_o^{\dagger}\partial_{x}\psi_o^{\phantom{\dagger}}\right).\label{H0evenodd}\end{equation}The open boundary conditions in the weak coupling limit, $\Psi(0)=\Psi(L)=0$, imply \begin{equation}
\psi_{e/o}(L)=\mp e^{-ik_FL}\psi_{e/o}(0).\label{eobc}
\end{equation} Because of the relative minus sign in Eq. (\ref{eobc}), the even and odd channels are nondegenerate.  The energy levels alternate between the even and channels. We assume that the lowest energy single-particle state in the band is even under parity. In this case, for $N_0=4p+1$, $p$ integer, the lowest unoccupied state  belongs to the even channel. The energy levels  relative to $\mu_0^*$ for $J=0$ are $\epsilon^e_n=(2n+1/2)\Delta$ and $\epsilon^o_n=(2n-1/2)\Delta$, with $\Delta=\pi v_F/L$ and $n$ integer.

We can account for the magnetic flux through the ring by introducing a phase in the hopping between the quantum dot and the two leads.\cite{persistentcurrent} This does not affect the Hamiltonian of the open chain in (\ref{H0evenodd}), but the Kondo interaction is modified to \begin{equation}H_K=J [\Psi^{\dagger}(\varepsilon)+e^{i\alpha}\Psi^{\dagger}(L-\varepsilon)]\frac{\vec{\sigma}}{2}[\Psi(\varepsilon)+e^{-i\alpha}\Psi(L-\varepsilon)]\cdot \vec{S},\end{equation}where $\alpha\equiv2\pi\Phi/\Phi_{0}$ and $\Phi_{0}=hc/e$ is the
flux quantum. The Kondo interaction in terms of even and odd fields reads
\begin{eqnarray}H_K&=&2\pi v_{F}\lambda_0\left[\cos\frac{\alpha}{2}\psi_{e}^{\dagger}\left(0\right)-i\sin\frac{\alpha}{2}\psi_{o}^{\dagger}\left(0\right)\right]\nonumber
 \\
 &  & \times\frac{\vec{\sigma}}{2}\left[\cos\frac{\alpha}{2}\psi_{e}^{\phantom{\dagger}}\left(0\right)+i\sin\frac{\alpha}{2}\psi_{o}^{\phantom{\dagger}}\left(0\right)\right]\cdot\vec{S},\label{eq:Kondowithflux}\end{eqnarray}
 where $\lambda_0=4J\sin^2(k_F\varepsilon)/\pi v_F$. If particle-hole symmetry is broken we have to add the potential scattering term\begin{eqnarray}H_p&=&2\pi v_F V\left[\cos\frac{\alpha}{2}\psi_{e}^{\dagger}\left(0\right)-i\sin\frac{\alpha}{2}\psi_{o}^{\dagger}\left(0\right)\right]\nonumber
 \\
 &  & \times\left[\cos\frac{\alpha}{2}\psi_{e}^{\phantom{\dagger}}\left(0\right)+i\sin\frac{\alpha}{2}\psi_{o}^{\phantom{\dagger}}\left(0\right)\right],\label{potential}\end{eqnarray} 
where $V\sim O(\lambda_0)$ is the dimensionless coupling constant. This operator is strictly marginal in the noninteracting case. We will assume $V\ll 1$ and treat $H_p$ using first order perturbation theory. 

For $\lambda=0$, the energy levels are flux-independent, due to the open boundary conditions. In the weak coupling limit 
$\lambda(L)\ll 1$, perturbation theory in the Kondo interaction of Eq.  (\ref{eq:Kondowithflux}) for general $\alpha$ yields a 
different splitting for even and odd channels. As a result, the spin part of the ground state energy depends on $\ell$ mod 4. 
To $O(\lambda_0^2)$, we find
\begin{eqnarray}
E(N_0+\ell)&=&E_0+\mu_0^*\ell+\Delta\left(\frac{\ell^2}{4}+s^2\right)\nonumber\\&&-3s^2\Delta\left(1-\cos 
\tilde\alpha_{\ell}\right)\lambda(L)\nonumber\\ & &-3s^2(\ln 2) \, \Delta \cos\tilde\alpha_{\ell}\left(1-\cos 
\tilde\alpha_{\ell}\right) \lambda^2(L)\nonumber\\& & +\frac{3\Delta }{16}\left(1+\cos^2\alpha\right)
 \lambda^2(L)+O\left(\lambda^3\right),\label{energyEQD}
\end{eqnarray}
where $\tilde\alpha_{\ell}\equiv \alpha+(\ell+1)\pi/2$; $s=0$ for $\ell$ even and $s=1/2$ for $\ell$ odd. In Eq. (\ref{energyEQD}) it is implicit that a $\lambda_0^2\ln L$ term is generated by perturbation theory with the correct coefficient to recover the expansion of $\lambda(L)$.   

The charge step locations can be calculated using Eq. (\ref{mulformula}). The last term in Eq. (\ref{energyEQD}) is independent 
of $\ell$ and drops out when taking the difference $E(N_0+\ell+1)-E(N_0+\ell)$.  To first order, the potential scattering term in
 Eq. (\ref{potential}) simply increases the  energy by $V\Delta  (1+\cos\alpha)$ for each electron added in the even channel 
and by $V\Delta (1-\cos\alpha)$ for each electron added in the odd channel. The charge steps in the weak coupling limit are 
located at (see dashed lines in  Fig.
\ref{cap:(Color-online)-Flux})\begin{eqnarray}
\frac{\mu_{\ell+\frac{1}{2}}-\mu_0^*}{\Delta}&=&\left\lfloor\frac{\ell}{2}\right\rfloor+\frac{1}{2}+\left[1+\cos\left(\alpha+\pi\left\lfloor\frac{\ell}{2}\right\rfloor\right)\right]\times\nonumber\\  & & \times \left[V-\frac{3}{4}(-1)^{\ell}\lambda(L)+O\left(\lambda^2\right)\right].\label{stepsfluxweak}
\end{eqnarray}
Note that, although potential scattering gives some weak flux dependence to the energy levels, it does not lift their spin degeneracy. Only the term due to the Kondo interaction alternates between $\ell$ even and $\ell$ odd and therefore lifts the spin degeneracy. Hence the main effect of the Kondo interaction in the weak coupling limit is to split the charge steps that are degenerate  for $\lambda=0$, namely $\mu_{\ell+1/2}$ and $\mu_{\ell+3/2}$ with $\ell$ even. We define \be \delta \mu_n\equiv\mu_{2n+\frac{3}{2}}-\mu_{2n+\frac{1}{2}},\ee where $n$ is an integer. From Eq. (\ref{stepsfluxweak}) we obtain \begin{equation}
\frac{\delta\mu_n(\alpha)}{2\Delta}=\frac{3}{4}\left[1+\left(-1\right)^{n}\cos\alpha\right]\lambda\left(L\right)+O\left(\lambda^{2}\right).\label{splitweak}\end{equation}
For symmetric coupling to the leads only the even channel couples to the impurity when $\alpha=2m\pi$, $m$ integer. As a result, the steps in the odd channel (corresponding to odd values of $n$ in Eq. (\ref{splitweak})) do not split. The opposite happens when $\alpha=(2m+1)\pi$ (see Fig.
\ref{cap:(Color-online)-Flux}).%

In the strong coupling limit, the $\pi/2$ phase shift modifies the
boundary conditions in such a way that the transmission across the
dot becomes perfect.\cite{delft,raikh,ng} For zero flux, the phase shift in the even channel implies \be
\psi_{e/o}(L)= e^{-ik_FL}\psi_{e/o}(0).\label{pbc_ring}\ee
With these new, periodic boundary conditions, the right- and left-moving components are decoupled\be
\psi_{R/L}(L)= e^{-ik_FL}\psi_{R/L}(0).
\ee The system is
then equivalent to an ideal 1D ring, whose eigenstates can be labeled as right or left movers. 
The degeneracy of right and left-moving energy levels at flux $m\pi$ implies charge steps of magnitude $4$ in 
the strong coupling limit, as can be seen  in Fig. \ref{cap:(Color-online)-Flux}. 
As we did in the weak coupling limit, we can introduce the flux in the hopping from $x=L$ to $x=0$. Since the ring is now closed, the unperturbed energy levels are flux dependent,\cite{fuhrer}\be
\epsilon_n^{R/L}(\alpha)=(2n-1/2\pm\alpha/\pi)\Delta,\label{levelssc}
\ee and
exhibit zigzag lines with level crossings at $\alpha=m\pi$, $m$ integer. The energy levels in Eq. (\ref{levelssc}) are measured 
from $\mu_0^*$ defined in the weak coupling limit. The shift of $-1/2$ is due to the phase shift of $\pi/2$ for 
the states in the even channel at $\alpha=0$ and can be understood in the following way. Consider the lattice model 
with $N-1$ sites and a symmetric Kondo coupling between the electrons at the ends of the chain, at $j=1$ and $j=N-1$, and 
 to the impurity spin at $j=0$,\bea
H & = & -t\sum_{j=1}^{L-2}\left(c_{j}^{\dagger}c_{j+1}^{\phantom{\dagger}}+h.c.\right)\nonumber\\
&&+J\left(c_1^\dagger+c^\dagger_{N-1}\right)\frac{\vec{\sigma}}{2}\left(c_1^{\phantom\dagger}+c^{\phantom\dagger}_{N-1}\right)\cdot \vec{S}.
\eea For $J=0$ and at half-filling, there are $N-1$ free conduction electrons in the chain. The hopping Hamiltonian is invariant under the particle-hole transformation \be
c_j\to (-1)^jc_j^\dagger.
\ee
For $J\neq 0$, particle-hole symmetry is broken by the Kondo interaction if $N$ is odd and preserved if $N$ is even, since under particle-hole transformation\be
c_1^{\phantom\dagger}+c^{\phantom\dagger}_{N-1}\to -\left[c_1^{\phantom\dagger}+(-1)^{N}c^{\phantom\dagger}_{N-1}\right].
\ee As we chose to fix $\mu_0^*$ so that $N_0=4p+1$ is odd, the charge steps in the strong coupling limit are not symmetric about $\mu_0^*$. However, the charge steps must be symmetric about $\mu_0^*\pm\Delta/2$, which correspond to values of chemical potential with an even number of electrons in the weak coupling limit. Notice that, if we start from $N_0$ odd in the weak coupling limit, there is a value of $\lambda(L)\sim O(1)$ (for $L\sim\xi_K$) at which the charge step $\mu_{\frac{1}{2}}$ crosses $\mu_0^*$ for $\alpha=0$. This means that, for fixed  $\mu=\mu_0^*$ and $\alpha=0$, the ground state at the strong coupling fixed point has an even number of  electrons ($N=N_0+1=4p+2$, i.e. $N_0-1=4p$ free electrons in the ring plus two in the singlet).

Using the strong coupling boundary conditions, the local Fermi liquid interaction analogous to Eq. (\ref{eq:localFLinteraction}) can be written\begin{eqnarray}
H_{FL}&=&-\frac{\pi^2}{6}\frac{v_F^2}{T_K}\left\{\left[\psi_R(0)+e^{-i\alpha}\psi_L(0)\right]^{\dagger}\right.\nonumber\\& &\qquad \left.\times\frac{\vec{\sigma}}{2}\left[\psi_R(0)+e^{-i\alpha}\psi_L(0)\right]\right\}^2.\label{FLring}
\end{eqnarray}
The main effect of this interaction is to lift the degeneracy between right and left movers at $\alpha=m\pi$ and to split two out of the four values
of $\mu_{\ell+\frac{1}{2}}$. We calculate the correction to the ground state energy using
degenerate perturbation theory to O($1/T_{K}$) for $V=0$. The calculation is similar to the weak coupling limit of the side-coupled quantum dot\cite{persistentcurrent} and is presented in detail in Appendix A. The charge steps (as labeled in the weak coupling limit) for $\alpha\approx m\pi$, $m$ even, are given by  (see Fig. \ref{cap:(Color-online)-Flux}). \begin{eqnarray}
\frac{\mu_{4n+1\pm{1 \over 2}}-\mu_0^*}{\Delta}&=&2n+{1\over 2}\pm1 \mp\frac{\pi\xi_K}{4L}\nonumber\\& &\mp\sqrt{\left(\frac{\alpha-m\pi}{\pi}\right)^2+\left(\frac{\pi\xi_K}{4L}\right)^2},\label{stepsci}\\
\frac{\mu_{4n+3\pm{1 \over 2}}-\mu_0^*}{\Delta}&=&2n+{3\over 2}\mp\sqrt{\left(\frac{\alpha-m\pi}{\pi}\right)^2+\left(\frac{\pi\xi_K}{4L}\right)^2}\nonumber\\& &\pm2\sqrt{\left(\frac{\alpha-m\pi}{\pi}\right)^2+\left(\frac{\pi\xi_K}{8L}\right)^2}.
\end{eqnarray}
For $\alpha\approx m\pi$, $m$ odd, the charge steps are located at \bea
\frac{\mu_{4n+1\pm{1 \over 2}}-\mu_0^*}{\Delta}&=&2n+{1\over 2}\mp\sqrt{\left(\frac{\alpha-m\pi}{\pi}\right)^2+\left(\frac{\pi\xi_K}{4L}\right)^2}\nonumber\\& &\pm2\sqrt{\left(\frac{\alpha-m\pi}{\pi}\right)^2+\left(\frac{\pi\xi_K}{8L}\right)^2},\\
\frac{\mu_{4n+3\pm{1 \over 2}}-\mu_0^*}{\Delta}&=&2n+{3\over 2}\pm1\mp\frac{\pi\xi_K}{4L}\nonumber\\& &\mp\sqrt{\left(\frac{\alpha-m\pi}{\pi}\right)^2+\left(\frac{\pi\xi_K}{4L}\right)^2}.\label{stepscf}
\eea
Note that two charge steps remain degenerate at each point of level crossing $\alpha= m\pi$. This is because for $\alpha=m\pi$ either the single-particle states in the odd channel (for $m$ even) or the states in the even channel (for $m$ odd) decouple from the spin of the dot. As a result, the Fermi liquid interaction reduces to  
\be
H_{FL}=-\frac{2\pi^2}{3}\frac{v_F^2}{T_K}\left\{\psi_{e,o}^\dagger(0)\frac{\vec\sigma}{2}\psi^{\phantom\dagger}_{e,o}(0)\right\}^2,
\ee
where the index $e$ applies to $m$ even and the index $o$ to $m$ odd. The two degenerate charge step correspond to adding electrons to the state whose energy is not modified by  the Fermi liquid interaction. 

If the  potential scattering is nonzero, the degeneracy between the even and odd channel at $\alpha=m\pi$ is lifted. For $\xi_K/L\sim V\ll1$, both the Fermi liquid interaction and the potential scattering can be treated using degenerate perturbation theory near $\alpha\approx m\pi$. If $\xi_K/L\ll V\sim O(1)$, the Fermi liquid interaction can be added on top of the unperturbed energy levels of an ideal ring with a delta function potential at $x=0$. In any case, the result is that, due to avoided level crossings, potential scattering suppresses the  flux dependence of the charge steps in the limit $\xi_K/L\ll1$. However, a feature that survives when $V\neq0$ is that the splitting  of the double steps in the even (odd) channel for $\alpha\approx m\pi$ with $m$ even ($m$ odd) is proportional to $\xi_K/L$.

\begin{figure}
\includegraphics[width=\columnwidth]{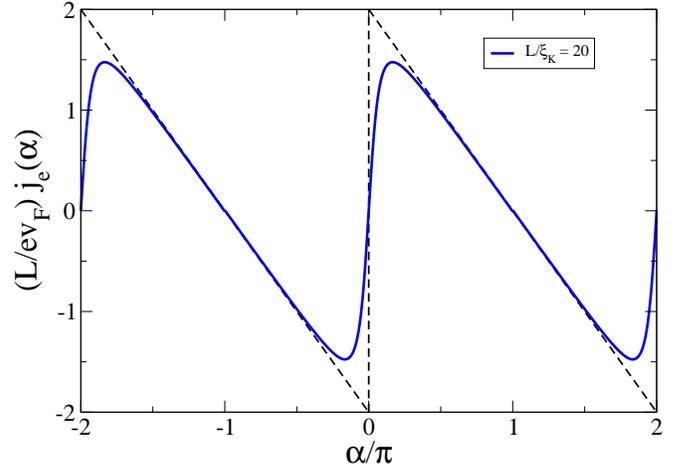}
\caption{(Color online) Persistent current for the embedded quantum dot with $N$ electrons, $N$ mod $4=0$, in the strong coupling limit. The dashed line represents the saw-tooth shape expected for periodic boundary conditions at the strong coupling fixed point. The solid line is the result of Eq. (\ref{jeven}) for $L/\xi_K=20$.  \label{pcurrent}}
\end{figure}

This significant difference in the flux dependence of the charge steps, between the weak and strong 
Kondo coupling limits, is consistent with previous results for the persistent current. 
The persistent current can be calculated by taking the derivative of the ground state energy with respect to the magnetic
 flux at fixed electron number: 
\be j=-e\,(\partial E/\partial \alpha)|_N ,\ee where $e$ is the electron charge. Using Eq. (\ref{energyEQD}), we recover the
 results derived in Ref. [\onlinecite{persistentcurrent}] for the weak coupling limit. Here we shall be concerned with 
the strong coupling limit. At the strong coupling fixed point, $\xi_K/L\rightarrow 0$, the persistent current has a
 saw-tooth shape as expected for an ideal ring (Fig. \ref{pcurrent}). For small but finite $\xi_K/L$,
 we expect that the Fermi liquid interaction rounds off the sharp features near the degeneracy points.

Consider first the case of $N$ odd. For both $N$ mod $4=1$ and $N$ mod $4=3$ the ground state is quasi-degenerate for $\alpha\approx m\pi$, $m$ integer. From first order perturbation theory in the Fermi liquid interaction, we find that the ground state energy for $\alpha\approx m\pi$ and $\xi_K/L\ll1$ is (see Appendix A)
\bea
\frac{E(\alpha)}{\Delta}&=&\frac{1}{2}\left(\frac{[\alpha]}{\pi}\right)^2+\frac{1}{2}\left(\frac{[\pi-\alpha]}{\pi}\right)^2+\frac{|\alpha-m\pi|}{\pi}-\frac{\pi\xi_K}{4L}\nonumber\\& &-\sqrt{\left(\frac{\alpha-m\pi}{\pi}\right)^2+\left(\frac{\pi\xi_K}{4L}\right)^2}+\textrm{const},\label{pcodd}
\eea
where $[\theta]$ denotes the value of $\theta$ reduced to the interval $-\pi<[\theta]\leq\pi$ by subtracting an integer multiple of $2\pi$, i.e., $[\theta]=\theta$ if $|\theta|<\pi$, $[\theta]=\theta-2\pi$ if $\pi<\theta<3\pi$, etc. The persistent current for $N$ odd and $\alpha\approx m\pi$ is\begin{equation}j_o(\alpha)=-\frac{ev_F}{L}\left[\frac{2(\alpha-m\pi)}{\pi}-\frac{(\alpha-m\pi)/\pi}{\sqrt{\left(\frac{\alpha-m\pi}{\pi}\right)^2+\left(\frac{\pi\xi_K}{4L}\right)^2}}\right].\label{PC_Nodd}\end{equation}
In the above case the persistent current is paramagnetic and has $\pi$ periodicity.\cite{persistentcurrent}

Now consider $N$ even. For $N$ mod $4=0$, the ground state is quasi-degenerate for $\alpha\approx m\pi$, $m$ even. The ground state energy near these points is \begin{eqnarray}\frac{E(\alpha)}{\Delta}&=&\left(\frac{[\pi-\alpha]}{\pi}\right)^2-\frac{\pi\xi_K}{4L}+\frac{2|\alpha-m\pi|}{\pi}\nonumber\\& &-2\sqrt{\left(\frac{\alpha-m\pi}{\pi}\right)^2+\left(\frac{\pi\xi_K}{8L}\right)^2}+\textrm{const}.\label{pceven1}\end{eqnarray}
For $N$ mod $4=2$, the degeneracy points appear at $\alpha\approx m\pi$, with $m$ odd. The ground state energy is \begin{eqnarray}\frac{E(\alpha)}{\Delta}&=&\left(\frac{[\alpha]}{\pi}\right)^2-\frac{\pi\xi_K}{4L}+\frac{2|\alpha-m\pi|}{\pi}\nonumber\\& &-2\sqrt{\left(\frac{\alpha-m\pi}{\pi}\right)^2+\left(\frac{\pi\xi_K}{8L}\right)^2}+\textrm{const}.\label{pceven2}\end{eqnarray}
Eq. (\ref{pceven2}) can be obtained by a translation $\alpha\to\alpha+\pi$ of Eq. (\ref{pceven1}). In both cases, the persistent current for $N$ even near $\alpha\approx m\pi$ ($m$ even for $N$ mod $4=2$ and $m$ odd for $N$ mod $4=2$) reads\be
j_e(\alpha)=-\frac{ev_F}{L}\left[\frac{2(\alpha-m\pi)}{\pi}-\frac{2(\alpha-m\pi)/\pi}{\sqrt{\left(\frac{\alpha-m\pi}{\pi}\right)^2+\left(\frac{\pi\xi_K}{8L}\right)^2}}\right]
\label{jeven}\ee
The persistent current  for $N$ even has $2\pi$ periodicity. It is paramagnetic for $N$ mod $4=0$ and diamagnetic for $N$ mod $4=2$. The persistent current for the embedded quantum dot with $N=4p$ in the limit $\xi_K\ll L$ is illustrated in Fig. \ref{pcurrent}.

The effect of electron-electron interactions on the transport through an embedded quantum dot was discussed in [\onlinecite{persistentcurrent,kim}]. For $K_c\approx1$, the system still flows to the one-channel Kondo fixed point, with one electron in the even channel screening the impurity spin. However, one important difference from the noninteracting case is that, for $K_c<1$, the backscattering term ($\sim \psi^\dagger_L(0)\psi_R(0)$) of $H_p$ in Eq. (\ref{potential}) is a relevant perturbation in the sense of the renormalization group. As in the Kane-Fisher problem,\cite{kanefisher} the effective potential scattering grows with $L$ as $V_{eff}\propto VL^{(1-K_c)/2}$ for $V_{eff}\ll D$ and as $V_{eff}\propto VL^{(1/K_c-1)/2}$ for $V_{eff}\gg D$. The large $V_{eff}$ suppresses the flux dependence of the charge steps in  the strong coupling limit of the Kondo effect and the persistent current decreases with $L$ faster than $1/L$.\cite{persistentcurrent}

\section{Experimental possibilities}

With current technology, it is not clear whether one can experimentally realize a ballistic single-channel ring of a suitable length to directly test the theoretical predictions for this geometry.  However, prospects are more encouraging for the linear geometry, where a straight wire segment is coupled to a localized spin at one end.  Candidate systems may include carbon nanotubes and wires made from semiconductor nanocrystals, as well one-dimensional conducting channels fabricated from two-dimensional semiconductor structures.~\cite{Biercuk2005, Hu2007, Jorgensen2007, Yacobi1996}
Perhaps the most promising candidates are one-dimensional wires constructed, using cleaved edge overgrowth, at the edge of a GaAs single-well or double-well structure.~\cite{Yacobi1996, Auslaender2000, Auslaender2002, Steinberg2006}   Electron mean-free paths of the order of 10 micrometers have been obtained in such structures, at electron densities in the range of 50 electrons per micron.  Local electron densities in the wire can be controlled by means of top gates, and one can thereby produce a series of conducting segments, of various lengths, separated by barriers of depleted regions.  Electrical contact to conducting wire segments can be made through a two-dimensional electron gas, which is present within the GaAs well in regions where it is not depleted by a top gate.  In double well structures, one can also establish momentum-controlled tunneling contacts between parallel wires in the two wells.~\cite{Auslaender2002, Steinberg2006}

In order to realize the geometry of interest here, one might create a short conducting segment with a small odd number of electrons
 (a Kondo dot), separated by a depleted barrier region from a much longer conducting segment, which would constitute the 
one-dimensional wire.  The strength of the Kondo coupling could be controlled by varying the voltage on the top gate above 
the barrier region.  Electron energy levels could be studied by tunneling electrons into the conducting segment from a lead at the 
opposite end from the Kondo dot, from a weakly coupled two-dimensional gas at the side, and/or by tunneling from a second parallel wire. 
  Tunneling measurements can give information about excited states of the wire system as well as the ground state.  Alternatively, the 
charge state of a conducting segment can be monitored through its Coulomb interaction with a nearby single electron transistor or
 quantum point contact device whose electrical conductance is sensitive to small changes in the electrostatic potential.~\cite{Hu2007}

Single-wall carbon nanotubes are another promising candidate for realizing effects analogous to those discussed in this paper. 
Electron mean-free-paths in nanotubes can be very long.  Gated quantum dots have been fabricated in carbon nanotubes,
~\cite{Biercuk2005, Jorgensen2007, Makarovski2007} and Kondo-type effects have been observed.~\cite{Makarovski2007} 
Analysis of effects in a carbon nanotube is made more complicated, however, by the existence of two orbital conducting channels,
 which are degenerate in an ideal nanotube, but may be split by distortion or imperfections in actual nanotubes.
\cite{Jorgensen2007, Makarovski2007}

\section{conclusion} \label{sec:conclusion}
We have shown that the width of the charge steps in a mesoscopic device depends on the finite size of the Kondo cloud. In general, the 
signature of this effect is the broadening of the Coulomb blockade valleys with total $N$ even as we increase $T_K$ (decrease $\xi_K$). 
For the case of a single-channel noninteracting wire, the crossover between weak ($L\ll\xi_{K}$) and strong ($L\gg\xi_{K}$) coupling is
 described by a universal scaling function $R(\xi_K/L)$, defined as the ratio between the width of the odd steps and the width of the 
even steps. Using perturbation theory in the Kondo interaction in the weak coupling limit and perturbation theory in the effective 
Fermi liquid interaction in the strong coupling limit, we derived the asymptotic behavior of $R(\xi_K/L)$ for $L\ll\xi_{K}$ and 
$L\gg\xi_{K}$. These formulas are in agreement with the exact numerical results obtained using the Bethe ansatz solution. We generalized 
these results for the case of a quantum dot coupled to a Luttinger liquid. If we neglect the effect of the marginal bulk interaction, the ratio $\tilde{R}$ between even and odd steps is
 still a scaling function of $\xi_K/L$, but the asymptotic values of $\tilde{R}$ for $L\ll\xi_K$ and $L\gg\xi_K$ are determined by the 
parameter $u=v_sK_c/v_c$. More generally, $\tilde{R}$ is also a function of the effective coupling constant $g(L)$ associated with the
 marginal bulk interaction. Finally, we have shown that, for the geometry of a quantum dot  embedded in a mesoscopic ring, the charge
 steps are weakly flux dependent for $L\ll\xi_K$ and strongly flux dependent for $L\gg\xi_K$. The latter behavior is reflected in the 
persistent current near the strong coupling fixed point, which we calculated using a local Fermi liquid theory.

\acknowledgements
We thank  J. A. Folk, C. M. Marcus, E. S\o rensen and A. Yacoby for useful discussions. We acknowledge support 
from CNPq-Brazil (RGP, 200612/2004-2), NSERC (RGP, NL, IA),  CIfAR (IA) and NSF (BIH, grant DMR-05-41988).

\appendix
\section{Charge steps for embedded quantum dot in the strong coupling limit}
In this appendix we show how the Fermi liquid interaction of Eq. (\ref{FLring}) lifts the fourfold degeneracy of the charge steps of the ideal ring near the points of level crossing. We focus on the calculation of ground state energy $E(N)$ for fixed $N$ and $\alpha\approx 0$ using degenerate perturbation theory. We generalize the results for $\alpha\approx m\pi$, $m$ integer, at the end of this section. 

\begin{figure}
\psfrag{epsL}{$\epsilon_n^L$}
\psfrag{epsR}{$\epsilon_n^R$}
\includegraphics[width=7cm,clip]{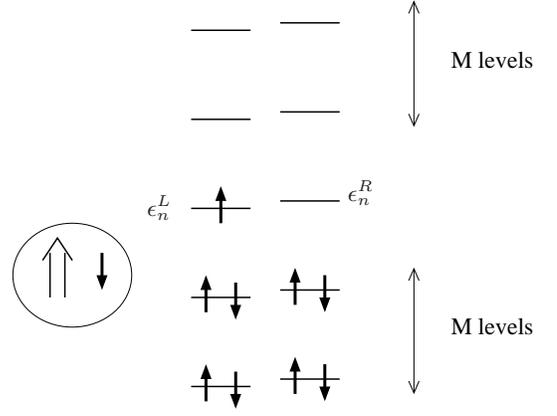}

\caption{Occupation of the energy levels of the embedded quantum dot in the strong coupling limit for $N=4(p+n)+3$ and $\alpha\approx0$. The pair of states with energy $\epsilon_n^{R}\approx\epsilon_n^L$ is partially occupied by a single electron.  \label{figlevels}}
\end{figure}

We start by separating the mode corresponding to the highest partially occupied energy level from the remaining ones that are either completely occupied  or completely empty. We write $N=N_0+\ell=4(p+n)+2+\ell^\prime$, with $n$ integer and $\ell^\prime\equiv(N-2)$ mod 4. Then the quasi-degenerate levels at the Fermi level are $\epsilon_{n}^{R/L}(\alpha)=2\Delta(n+3/4\pm\alpha/2\pi)>0$ for $N>N_0+1$  (see Fig. \ref{figlevels}). (Here we measure the energy levels from $\mu_0^*$ and set $\mu_0^*=0$ for a shorthand notation.) We rewrite the Fermi liquid interaction of Eq. (\ref{FLring}) in the form\be
H_{FL}=-\frac{\Delta}{6}\frac{\pi \xi_K}{L}\left\{(\chi_0+\chi^\prime)^\dagger\frac{\vec{\sigma}}{2}(\chi_0+\chi^\prime)\right\}^2,
\ee
where $\chi_0=c_{nR}+e^{-i\alpha}c_{nL}$, with $c_{nR/L}$ the operators that annihilates electrons in the states with energy $\epsilon_n^{R/L}$. The other modes are contained in $\chi^\prime = \sum_{n^\prime\neq n}(c_{n^\prime R}+e^{-i\alpha}c_{n^\prime L})$, with the sum restricted to states near the Fermi surface, i.e. $|\epsilon_{n^\prime}^{R/L}|\ll D $, where $D\ll \epsilon_F$ is the cutoff. We expand the interaction into  four terms
\be
\delta H_{FL}=\delta H_{FL}^{(1)}+\delta H_{FL}^{(2)}+\delta H_{FL}^{(3)}+\delta H_{FL}^{(4)}.
\ee
The first term, \be
H_{FL}^{(1)}=-\frac{\Delta}{6}\frac{\pi \xi_K}{L}\left(\chi_0^\dagger\frac{\vec{\sigma}}{2}\chi_0\right)^2,
\ee only involves $\chi_0$ and  couples quasi-degenerate states with the same total number $N$ which differ by the distribution of electrons in the partially filled level. Before we look at the effects of this term, we argue that the other three terms can be neglected. The second term is \bea
\delta H_{FL}^{(2)}&=&-\frac{\Delta}{6}\frac{\pi \xi_K}{L}\chi^{\prime\dagger}\frac{\vec{\sigma}}{2}\chi^\prime\cdot\chi^{\prime\dagger}\frac{\vec{\sigma}}{2}\chi^\prime,
\eea
This operator is diagonal in the subspace of states with fixed $N$ since it does not contain $\chi_0$ and does not act on the electrons occupying the partially filled level. The contribution to the ground state energy is
\bea
\delta E_{FL}^{(2)}&=&-\frac{\Delta}{6}\frac{\pi \xi_K}{L}\left\langle\chi^{\prime\dagger}\frac{\vec{\sigma}}{2}\chi^\prime\cdot\chi^{\prime\dagger}\frac{\vec{\sigma}}{2}\chi^\prime\right\rangle,
\eea
where $\left\langle \right\rangle$ denotes the expectation value in one of the states for  fixed $N$.  We obtain
\bea
\delta E_{FL}^{(2)}& =& -\frac{\Delta}{6}\frac{\pi \xi_K}{4L}(\sigma^a)_\mu^\nu(\sigma_a)_\lambda^\rho\sum_{n_1^\prime,m_1^\prime}\sum_{n_2^\prime,m_2^\prime} \nonumber\\ & &  \times \left\langle (c_{n_1^\prime R}^{\mu\dagger} +e^{i\alpha}c_{n_1^\prime L}^{\mu\dagger} )(c^{\phantom\dagger}_{m_1^\prime R\nu } +e^{-i\alpha}c^{\phantom\dagger}_{m_1^\prime L\nu } )\right.\nonumber\\&&\times \left.(c_{n_2^\prime R}^{\lambda\dagger} +e^{i\alpha}c_{n_2^\prime L}^{\lambda\dagger} )(c^{\phantom\dagger}_{m_2^\prime R\rho } +e^{-i\alpha}c^{\phantom\dagger}_{m_2^\prime L\rho } )\right\rangle\nonumber\\
& =&-\frac{\Delta}{6}\frac{\pi \xi_K}{L}(\vec{\sigma}^2)_\mu^\mu \sum_{n_1=n-M}^{n-1}\sum_{m_1^\prime=n+1}^{n+M} 1\nonumber\\& =&- \frac{3D^2}{2T_K},\label{cutoffdepterm}
\eea
where $M=DL/(2\pi v_F)$ is the number of states above or below $\epsilon_n$ and inside the cutoff. We choose the cutoff  to be symmetric without loss of generality, since the effects of particle-hole symmetry breaking can be cast into the potential scattering term of Eq. (\ref{potential}). The important point is that the  $O(1)$ contribution to $E(N)$ in Eq. (\ref{cutoffdepterm}) is cutoff dependent, but is independent of the occupation of the levels  $\epsilon^{R/L}_n$ (independent of $\ell^\prime$). Therefore, it gets cancelled when we take the difference $E(N+1)-E(N)$ and has no effect on the charge steps.

The last two terms are \bea
\delta E_{FL}^{(3)}&=&-\frac{\Delta}{6}\frac{\pi \xi_K}{L}\left\langle\chi_0^{\dagger}\frac{\vec{\sigma}}{2}\chi^\prime\cdot\chi^{\prime\dagger}\frac{\vec{\sigma}}{2}\chi_0\right\rangle \\ 
\delta E_{FL}^{(4)}& =& -\frac{\Delta}{6}\frac{\pi \xi_K}{L}\left\langle\chi^{\prime\dagger}\frac{\vec{\sigma}}{2}\chi_0\cdot\chi_0^{\dagger}\frac{\vec{\sigma}}{2}\chi^\prime\right\rangle.
\eea
In principle,  $\left\langle \right\rangle$ should be regarded as a matrix in the subspace of quasi-degenerate states for fixed $N$. Let $\left|a\right\rangle \otimes\left|FS\right\rangle$ and $\left|b\right\rangle\otimes\left|FS\right\rangle$ denote two states in this subspace, with $\left|a\right\rangle,\left|b\right\rangle$ the states of the electrons in the partially filled level and $\left|FS\right\rangle\equiv\prod_{n^\prime<n,\sigma=\uparrow,\downarrow}c^{\sigma\dagger}_{n^\prime R}c^{\sigma\dagger}_{n^\prime L}\left|0\right\rangle$ the filled Fermi sea. The corresponding matrix element is \be
\delta E_{FL}^{(3)}=-\frac{\Delta}{6}\frac{\pi \xi_K}{4L}(\sigma^a)_\mu^\nu(\sigma_a)_\lambda^\rho M\delta_\nu^\lambda\left\langle a\left|\chi_0^{\mu\dagger}\chi^{\phantom\dagger}_{0\rho}\right|b\right\rangle.\label{EFL3}
\ee
Likewise,
\be
\delta E_{FL}^{(4)}=-\frac{\Delta}{6}\frac{\pi \xi_K}{4L}(\sigma^a)_\mu^\nu(\sigma_a)_\lambda^\rho M\delta_\rho^\mu\left\langle a\left|\chi^{\phantom\dagger}_{0\nu}\chi_0^{\lambda\dagger}\right|b\right\rangle.\label{EFL4}
\ee
Note that the factor of $M$ is the same in Eqs. (\ref{EFL3}) and  (\ref{EFL4}) because the cutoff is particle-hole symmetric. Combining (\ref{EFL3}) and  (\ref{EFL4}), we get\bea
\delta E_{FL}^{(3)}+\delta E_{FL}^{(4)}&=&-\frac{\Delta}{2}\frac{\pi\xi_K}{4L}\frac{DL}{2\pi v_F} \left\langle a\left|\left\{\chi^{\phantom\dagger}_{0\mu},\chi_0^{\mu\dagger}\right\}\right|b\right\rangle\nonumber\\&=& \frac{\pi\xi_K}{4L} D\delta^{ab}.
\eea
As a result,  this contribution is also diagonal in the subspace of fixed $N$ and is independent of $\ell^\prime$.  Therefore, to first order in $\xi_K/L$, the splitting of the charge steps is determined by $H_{FL}^{(1)}$ only.

We now turn to the calculation of $\delta E_{FL}^{(1)}$ using degenerate perturbation theory.
The trivial case is the ground state energy for $N$ mod $4=2$ ($\ell=4n+1$, $n$ integer). In this case, which includes the ground state for $\mu=\mu_0^*$ at $\alpha\approx0$, the ground state is non-degenerate: two electrons are bound in the singlet and the remaining $4(p+n)$ electrons fill up the single-particles energy levels of the ideal ring.  This situation corresponds to $\ell^\prime=0$, and the contribution from $\delta E_{FL}^{(1)}$ vanishes:
\be
\delta E_{FL}^{(1)}[N=4(p+n)+2]=0.\label{Ea01}
\ee

Now consider $N$ mod $4=3$ ($\ell^\prime=1$). For $\xi_K/L=0$, there is one extra electron on top of the filled Fermi sea (as illustrated in Fig. \ref{figlevels}). Since the Fermi liquid interaction commutes with the total spin, we can work in the subspace where this extra electron has spin up. The two quasi-degenerate states the electron can occupy are \be
\left|1\right\rangle=c^{\uparrow\dagger}_{nR}\left|FS\right\rangle\quad,\quad
\left|2\right\rangle=c^{\uparrow\dagger}_{nL}\left|FS\right\rangle,
\ee where $\left|FS\right\rangle\equiv\left|GS(N=4(p+n)+2)\right\rangle$ is the filled Fermi sea. The unperturbed energies (for $\xi_K/L=0$) are\bea
E_{R}&=&\left\langle1\left|H_0\right|1\right\rangle=E[4(p+n)+2]+\epsilon_n^{R},\\
E_{L}&=&\left\langle2\left|H_0\right|2\right\rangle=E[4(p+n)+2]+\epsilon_n^{L}.
\eea
It is convenient to rewrite $H_{FL}^{(1)}$ in the form\bea
H_{FL}^{(1)}&=&-\frac{\Delta}{6}\frac{\pi\xi_K}{L}\left\{ (c^\dagger_{nR}+e^{i\alpha}c^\dagger_{nL})\frac{\vec{\sigma}}{2}(c^{\phantom\dagger}_{nR}+e^{-i\alpha}c^{\phantom\dagger}_{nL}) \right\}^2\nonumber\\
&=&-\frac{\Delta}{6}\frac{\pi\xi_K}{L}\left\{ \left(\vec{s}_{nR}+\vec{s}_{nL}\right)^2 \right.\nonumber\\
& &+c^\dagger_{nR}\frac{\vec{\sigma}}{2}c^{\phantom\dagger}_{nL}\cdot c^\dagger_{nL}\frac{\vec{\sigma}}{2}c^{\phantom\dagger}_{nR}+c^\dagger_{nL}\frac{\vec{\sigma}}{2}c^{\phantom\dagger}_{nR}\cdot c^\dagger_{nR}\frac{\vec{\sigma}}{2}c^{\phantom\dagger}_{nL}\nonumber\\
&&+2(\vec{s}_{nR}+\vec{s}_{nL})\cdot\left(e^{-i\alpha}c^\dagger_{nR}\frac{\vec{\sigma}}{2}c^{\phantom\dagger}_{nL}+h.c.\right)\nonumber\\&&\left.
+\left(e^{-i2\alpha}c^\dagger_{nR}\frac{\vec{\sigma}}{2}c^{\phantom\dagger}_{nL}\cdot c^\dagger_{nR}\frac{\vec{\sigma}}{2}c^{\phantom\dagger}_{nL}+h.c.\right)\right\},\label{hfl1}
\eea
where $\vec{s}_{nR/L}=c^\dagger_{nR/L}(\vec{\sigma}/2)c^{\phantom\dagger}_{nR/L}$.We calculate the matrix elements of $H_{FL}^{(1)}$ in the subspace of states $\left|1\right\rangle,\left|2\right\rangle$.  The associated matrix for the Hamiltonian including the Fermi liquid interaction is\be
\left\langle H_0+H_{FL}^{(1)}\right\rangle=\left( \begin{array}{cc}
E_R -\frac{\Delta\pi\xi_K}{4L}& -\frac{\Delta\pi\xi_K}{4L} e^{-i\alpha} \\
-\frac{\Delta\pi\xi_K}{4L} e^{i\alpha} & E_L -\frac{\Delta\pi\xi_K}{4L} \end{array} \right)
\ee Diagonalizing this matrix, we find that the ground state energy for $\ell^\prime=1$ is\bea
&&E[N=4(p+n)+3]\nonumber\\&=&E[N=4(p+n)+2]+\frac{\epsilon_n^R +\epsilon_n^L}{2}\nonumber\\&&-\frac{\Delta\pi\xi_K}{4L}-\sqrt{\left(\frac{\epsilon_n^R-\epsilon_n^L}{2}\right)^2+\left(\frac{\Delta\pi\xi_K}{4L}\right)^2}\nonumber\\
&=&E[N=4(p+n)+2]+\Delta\left(2n+\frac{3}{2}\right)\nonumber\\&&-\frac{\Delta\pi\xi_K}{4L}-\Delta\sqrt{\left(\frac{\alpha}{\pi}\right)^2+\left(\frac{\pi\xi_K}{4L}\right)^2}.\label{Ea02}
\eea 

For $N$ mod $4=0$ ($\ell^\prime=2$), we have to distribute two extra electrons in the levels with energies $\epsilon^{R/L}_n$. For $\xi_K/L=0$, the subspace with total $s_n^z=s_{nR}^{z}+s_{nL}^z=0$  is spanned by four states
\bea
\left|RR\right\rangle&=&c^{\downarrow\dagger}_{nR}c^{\uparrow\dagger}_{nR}\left|FS\right\rangle,\nonumber\\
\left|LL\right\rangle&=&c^{\downarrow\dagger}_{nL}c^{\uparrow\dagger}_{nL}\left|FS\right\rangle,\nonumber\\
\left|RL,s\right\rangle&=&\frac{1}{\sqrt{2}}\left(c^{\downarrow\dagger}_{nR}c^{\uparrow\dagger}_{nL}-c^{\uparrow\dagger}_{nR}c^{\downarrow\dagger}_{nL}\right)\left|FS\right\rangle,\nonumber\\
\left|RL,t\right\rangle&=&\frac{1}{\sqrt{2}}\left(c^{\downarrow\dagger}_{nR}c^{\uparrow\dagger}_{nL}+c^{\uparrow\dagger}_{nR}c^{\downarrow\dagger}_{nL}\right)\left|FS\right\rangle .\nonumber
\eea
Of the above states, the first three are singlets (total $s_n=0$) and the last one is a triplet ($s_n=1$). The corresponding unperturbed energies are \bea
E_{RR}&=&E[N=4(p+n)+2]+2\epsilon_n^{R},\nonumber\\
E_{LL}&=&E[N=4(p+n)+2]+2\epsilon_n^{L},\nonumber\\
E_{RL,s}&=&E_{RL,t}=E[N=4(p+n)+2]+\epsilon_n^{R}+\epsilon_n^{L}.\nonumber\eea There are still two other states with $s_n^z=\pm1$ that are quasi-degenerate with the above states. 
For $\xi_K/L\neq0$, the Fermi liquid interaction can only mix the states in the $s_n^z=0$ subspace. We note that $\left|RL,t\right\rangle$ does not couple to the other three states because $H_{FL}^{(1)}$ commutes with the total spin. Moreover, it is apparent that $\left|RL,s\right\rangle$ can only couple to $\left|RR\right\rangle$ or $\left|LL\right\rangle$ via the term of Eq. (\ref{hfl1}) that transfers one electron between right and left channels, \be
\sim (\vec{s}_{nR}+\vec{s}_{nL})\cdot\left(e^{-i\alpha}c^\dagger_{nR}\frac{\vec{\sigma}}{2}c^{\phantom\dagger}_{nL}+h.c.\right).
\ee
However, this term involves the total spin operator $\vec{s}_n=\vec{s}_{nR}+\vec{s}_{nL}$, which annihilates singlet states. Therefore, $\left|RL,s\right\rangle$ does not couple to $\left|RR\right\rangle$ or $\left|LL\right\rangle$ either. As a result, we just have  to diagonalize the $2\times2$ matrix spanned by $\{\left|RR\right\rangle,\left|LL\right\rangle\}$
\be
\left\langle H_0+H_{FL}^{(1)}\right\rangle=\left( \begin{array}{cc}
E_{RR} -\frac{\Delta\pi\xi_K}{4L}& -\frac{\Delta\pi\xi_K}{4L} e^{-i\alpha} \\
-\frac{\Delta\pi\xi_K}{4L} e^{i\alpha} & E_{LL} -\frac{\Delta\pi\xi_K}{4L}\end{array} \right).\label{matrix2}
\ee
The smallest eigenvalue of (\ref{matrix2}) yields \bea
&&E[N=4(p+n)+4]\nonumber\\&=&E[N=4(p+n)+2]+\epsilon_n^R +\epsilon_n^L\nonumber\\&&-\frac{\Delta\pi\xi_K}{4L}-\sqrt{\left(\epsilon_n^R-\epsilon_n^L\right)^2+\left(\frac{\Delta\pi\xi_K}{4L}\right)^2}\nonumber\\
&=&E[N=4(p+n)+2]+2\Delta \left(2n+\frac{3}{2}\right)\nonumber\\&&-\frac{\Delta\pi\xi_K}{4L}-\Delta\sqrt{\left(\frac{2\alpha}{\pi}\right)^2+\left(\frac{\pi\xi_K}{4L}\right)^2}.\label{e04n}
\eea 
The corrections to $E_{RL,s}$ and $E_{RL,t}$ at first order in $\xi_K/L$ are given by\bea
\left\langle RL,s\left|\delta H_{FL}^{(1)}\right|RL,s\right\rangle&=& 0,\\
\left\langle RL,t\left|\delta H_{FL}^{(1)}\right|RL,t\right\rangle&=&-\frac{\Delta\pi\xi_K}{2L}.
\eea
The ground state energy given in Eq.  (\ref{e04n}) is lower than $E_{RL,s},E_{RL,t}$, except at $\alpha=0$, where it is equal to $E_{RL,t}$. In fact, for $\alpha=0$, we have $\epsilon_n^R=\epsilon_n^L$. It is convenient to use the basis of even and odd channels, in which the Fermi liquid interaction of Eq. (\ref{hfl1}) assumes the form \be
H_{FL}^{(1)}=-\frac{2\Delta}{3}\frac{\pi\xi_K}{L}\vec{s}_{ne}^2,
\ee 
where $\vec{s}_{ne}=(c^\dagger_{nR}+c^\dagger_{nL})(\vec{\sigma}/4)(c^{\phantom\dagger}_{nR}+c^{\phantom\dagger}_{nL})$ is the spin operator for electrons in the even channel. The ground state for $\ell^\prime=2$ is obtained by adding one electron to the even channel  and one electron to the odd channel. In this case, $s_{ne}=1/2$ and $\delta E=-\Delta \pi\xi_K/(2L)$. For this given energy there are two degenerate states with $s_n^z=0$:\bea
\left|eo\right\rangle&=&c^{\downarrow\dagger}_{ne}c^{\uparrow\dagger}_{no}\left|FS\right\rangle=\frac{c^{\downarrow\dagger}_{nR}+c^{\downarrow \dagger}_{nL}}{\sqrt{2}}\frac{c^{\uparrow\dagger}_{nR}-c^{\uparrow \dagger}_{nL}}{\sqrt{2}}\left|FS\right\rangle,\nonumber\\
\left|oe\right\rangle&=&c^{\downarrow\dagger}_{no}c^{\uparrow\dagger}_{ne}\left|FS\right\rangle=\frac{c^{\downarrow\dagger}_{nR}-c^{\downarrow \dagger}_{nL}}{\sqrt{2}}\frac{c^{\uparrow\dagger}_{nR}+c^{\uparrow \dagger}_{nL}}{\sqrt{2}}\left|FS\right\rangle.\nonumber\\
\eea
These can be recognized as linear combinations of $\left|RL,t\right\rangle$ and $(\left|RR\right\rangle+\left|LL\right\rangle)/\sqrt{2}$. The latter is the eigenstate of the matrix in Eq. (\ref{matrix2}) with eigenvalue given by Eq. (\ref{e04n}) for $\alpha=0$. One can also verify that the energy of the states with $s_n^z=\pm1$ is lowered by the same amount as $E_{RL,t}$. For $\alpha=0$, this corresponds to putting one electron in the even channel and the other in odd channel, both with spin up or both with spin down.

Finally, the calculation of the ground state energy for $N$ mod $4=1$ ($\ell^\prime=3$) is analogous to the one for $N$ mod $4=3$ ($\ell^\prime=1$). We find\bea
&&E[N=4(p+n)+5]\nonumber\\&=&E[N=4(p+n)+2]+\frac{3(\epsilon_n^R +\epsilon_n^L)}{2}\nonumber\\&&-\frac{\Delta\pi\xi_K}{4L}-\sqrt{\left(\frac{\epsilon_n^R-\epsilon_n^L}{2}\right)^2+\left(\frac{\Delta\pi\xi_K}{4L}\right)^2}\nonumber\\
&=&E[N=4(p+n)+2]+3\Delta \left(2n+\frac{3}{2}\right)\nonumber\\&&-\frac{\Delta\pi\xi_K}{4L}-\Delta\sqrt{\left(\frac{\alpha}{\pi}\right)^2+\left(\frac{\pi\xi_K}{4L}\right)^2}.\label{Ea03}
\eea 

Now consider $\alpha\approx\pi$. In this case we have to consider that the level crossing involves the unperturbed levels $\epsilon_{n-1}^R(\alpha\approx \pi)\approx \epsilon_{n}^L(\alpha\approx \pi)\approx 2\Delta(n+1/4)$. The trivial case of filled shells, in which the correction to the ground state energy due to the Fermi liquid interaction vanishes,  occurs for $N$ mod $4=0$. In analogy with the calculation for $\alpha\approx0$, we find that for $\alpha\approx \pi$ \bea
&&E[N=4(p+n)+1]\nonumber\\&=&E[N=4(p+n)]+\Delta\left(2n+\frac{1}{2}\right)\nonumber\\&&-\frac{\Delta\pi\xi_K}{4L}-\Delta\sqrt{\left(\frac{\alpha-\pi}{\pi}\right)^2+\left(\frac{\pi\xi_K}{4L}\right)^2},\label{Epi1}
\eea
\bea
&&E[N=4(p+n)+2]\nonumber\\&=&E[N=4(p+n)]+2\Delta\left(2n+\frac{1}{2}\right)\nonumber\\&&-\frac{\Delta\pi\xi_K}{4L}-\Delta\sqrt{\left[\frac{2(\alpha-\pi)}{\pi}\right]^2+\left(\frac{\pi\xi_K}{4L}\right)^2},\label{Epi2}
\eea 
\bea
&&E[N=4(p+n)+3]\nonumber\\&=&E[N=4(p+n)]+3\Delta\left(2n+\frac{1}{2}\right)\nonumber\\&&-\frac{\Delta\pi\xi_K}{4L}-\Delta\sqrt{\left(\frac{\alpha-\pi}{\pi}\right)^2+\left(\frac{\pi\xi_K}{4L}\right)^2}.\label{Epi3}
\eea

The results in Eqs. (\ref{Ea01}), (\ref{Ea02}), (\ref{e04n}) and (\ref{Ea03}) apply directly to the case $\alpha\approx m\pi$, $m$ even,  if $\alpha$ is replaced by $\alpha-m\pi\ll1$. The results in Eqs (\ref{Epi1}-\ref{Epi3}) apply to the case  $\alpha\approx m\pi$, $m$ odd, if we replace $\alpha-\pi$ by $\alpha-m\pi$.
The expressions for the charge steps in Eqs. (\ref{stepsci} - \ref{stepscf}) are obtained by taking the difference $\mu_{\ell+1/2}=E(N_0+\ell+1)-E(N_0+\ell)=E(4p+\ell+2)-E(4p+\ell+1)$. 

Note also that for $\xi_K/L\to0$ the expressions for $E(N)$ and the charge steps reduce to the exact ones for the strong coupling fixed point for all values of $\alpha$. Defining $E^{(0)}(\alpha)$ as the ground state energy for $\xi_K/L=0$, the ground state energy for $\xi_K/L\ll1$ and $\alpha\approx m\pi$ can be written as\bea
E(N\, \textrm{odd})&=&E^{(0)}(N\, \textrm{odd})+\Delta\frac{|[\alpha-m\pi]|}{\pi}-\frac{\Delta\pi\xi_K}{4L}\nonumber\\&&-\Delta\sqrt{\left(\frac{[\alpha-m\pi]}{\pi}\right)^2+\left(\frac{\pi\xi_K}{4L}\right)^2}.\\
E(N\, \textrm{even})&=&E^{(0)}(N\, \textrm{even})+2\Delta\frac{|[\alpha-m\pi]|}{\pi}-\frac{\Delta\pi\xi_K}{4L}\nonumber\\&&-\Delta\sqrt{\left(\frac{2[\alpha-m\pi]}{\pi}\right)^2+\left(\frac{\pi\xi_K}{4L}\right)^2}.
\eea
These are the expressions used in Eqs. (\ref{pcodd}) and (\ref{pceven1}) to calculate the persistent current.

\end{document}